\documentclass[sigplan, 10pt]{acmart}

\makeatletter
\def\@ACM@checkaffil{}
\makeatother

\acmYear{2024}\copyrightyear{2024}
\setcopyright{acmlicensed}
\acmConference[EuroSys '24]{European Conference on Computer Systems}{April 22--25, 2024}{Athens, Greece}
\acmBooktitle{European Conference on Computer Systems (EuroSys '24), April 22--25, 2024, Athens, Greece}
\acmPrice{15.00}
\acmDOI{10.1145/3627703.3629580}
\acmISBN{979-8-4007-0437-6/24/04}

\usepackage[english]{babel}
\usepackage{blindtext}
\usepackage{textcomp}
\usepackage{makecell}
\usepackage{graphicx}
\usepackage{comment}
\usepackage{multirow}
\usepackage{fancyhdr}
\usepackage{xspace}
\usepackage{xcolor}
\usepackage{dsfont}
\usepackage{pgfplots}
\usepackage{multirow}
\usepackage{float}
\usepackage{microtype}
\usepackage{subfig}
\usepackage{booktabs}
\usepackage{hyperref}
\usepackage{algorithmic}
\usepackage{bussproofs}
\usepackage{mathtools}

\pgfplotsset{compat = 1.17}
\usetikzlibrary{patterns}
\usetikzlibrary{calc}
\pgfplotsset{every tick label/.append style={font=\small}}

\mathtoolsset{centercolon}

\newcommand{\hoare}[3]{\{\,#1\,\}\;#2\;\{\,#3\,\}}

\newenvironment{scprooftree}[1]%
  {\gdef\scalefactor{#1}\begin{center}\proofSkipAmount \leavevmode}%
  {\scalebox{\scalefactor}{\DisplayProof}\proofSkipAmount \end{center} }

\NewCommandCopy{\oldtexttt}{\texttt}
\renewcommand\texttt[1]{\oldtexttt{\fontsize{9.5pt}{11pt}\selectfont#1}}

\makeatletter
\newcolumntype{"}{@{\hskip\tabcolsep\vrule width 1pt\hskip\tabcolsep}}
\makeatother

\def \OurSystem{\textit{HAP}\xspace}

\title{\OurSystem: SPMD DNN Training on Heterogeneous GPU Clusters with Automated Program Synthesis}

\author{Shiwei Zhang}
\affiliation{%
  \institution{The University of Hong Kong}
}
\email{swzhang@cs.hku.hk}

\author{Lansong Diao}
\affiliation{%
  \institution{Alibaba Group}
}
\email{lansong.dls@alibaba-inc.com}

\author{Chuan Wu}
\affiliation{%
  \institution{The University of Hong Kong}
}
\email{cwu@cs.hku.hk}

\author{Zongyan Cao}
\affiliation{%
  \institution{Alibaba Group}
}
\email{zongyan.cao@alibaba-inc.com}

\author{Siyu Wang}
\affiliation{%
  \institution{Alibaba Group}
}
\email{siyu.wsy@alibaba-inc.com}

\author{Wei Lin}
\affiliation{%
  \institution{Alibaba Group}
}
\email{weilin.lw@alibaba-inc.com}



\begin{document}

\begin{abstract}
Single-Program-Multiple-Data (SPMD) parallelism has recently been adopted to train large deep neural networks (DNNs).
Few studies have explored its applicability on heterogeneous clusters, to fully exploit available resources for large
model learning. This paper presents \OurSystem, an automated system designed to expedite SPMD DNN training on
heterogeneous clusters. \OurSystem jointly optimizes the tensor sharding strategy, sharding ratios across heterogeneous
devices and the communication methods for tensor exchanges for optimized distributed training with SPMD parallelism. We
novelly formulate model partitioning as a program synthesis problem, in which we generate a distributed program from
scratch on a distributed instruction set that semantically resembles the program designed for a single device, and
systematically explore the solution space with an A*-based search algorithm. We derive the optimal tensor sharding
ratios by formulating it as a linear programming problem. Additionally, \OurSystem explores tensor communication
optimization in a heterogeneous cluster and integrates it as part of the program synthesis process, for automatically
choosing optimal collective communication primitives and applying sufficient factor broadcasting technique. Extensive
experiments on representative workloads demonstrate that \OurSystem achieves up to 2.41x speed-up on heterogeneous
clusters.
\end{abstract}

\maketitle

\section{Introduction}

Recent machine learning research has demonstrated that scaling up deep neural network (DNN) models not only improves
their prediction performance but also expands their capabilities. For instance, a language model can perform a task with
few-shot prompting after reaching a certain model scale \cite{emergent}. Massive models with billions or trillions of
parameters have emerged \cite{gpt3,palme}. Training these models necessitates the use of large clusters of accelerator
devices (dominantly GPUs), as well as sophisticated parallelization paradigms. Consequently, a number of parallelization
schemes have been proposed and adopted in distributed training of large DNNs, including data/model/pipeline parallelism
\cite{tofu,hypar,deepspeed,flexflow,unity,gspmd,hidup,pipedream,gpipe}. However, the majority of the existing proposals
concentrate on DNN training on homogeneous clusters, where all devices are of the same type and interconnect network
links have identical bandwidth.

On the other hand, the rapid evolution of accelerate devices (e.g., GPUs, TPUs, Habana chips) and multi-tenant resource
sharing have resulted in mixed device types and uneven interconnect bandwidth in many clusters. Efficient exploitation
of available heterogeneous resources for DNN training has piqued significant interest from AI practitioners
\cite{heterog,hdp,tag,gavel,accpar,prague,giph}.
A prevalent approach for making use of heterogeneous clusters is to distribute multiple training jobs onto different
homogeneous subsets of the cluster \cite{hived,mlaas}. However, this approach imposes a constraint on the maximum model
size, as it is bound by the capacity of the subsets. With the emergence of large language models, the demand for
employing the whole heterogeneous cluster to train a large model has become increasingly crucial.

The limited existing heterogeneity-aware DNN training designs mostly support data parallelism and inter-op model
parallelism \cite{heterog,hdp,tag}. For data parallelism, each device trains the DNN model for a distinct portion of the
dataset, allocated according to capacity of the device, and synchronizes model gradients among each other at the end of
each training iteration. Inter-op model parallelism involves placing different operations in the DNN model across
devices, with intermediate tensors transmitted among devices during training. For better GPU utilization,
inter-op model-parallel training is often scheduled in a pipelined manner, with micro-batches of data processed on
different devices at the same time. Due to constrained memory capacity of GPUs, pure data parallelism or inter-op model
parallelism may not be feasible or efficient for training large DNN models. For instance, in models with
Mixture-of-Expert (MoE) \cite{moe} layers, a single tensor may surpass the GPU's memory limit. In these cases, it
becomes necessary to employ intra-op model parallelism, which partitions the operations/tensors and distributes the
shards to different devices. It is especially challenging to devise efficient strategies on tensor sharding and
deployment across a cluster of heterogeneous devices, due to the large strategy space.

This work studies Single-Program-Multiple-Data (SPMD) parallelism for efficient training of large DNNs on heterogeneous
clusters. SPMD parallelism generalizes data parallelism and intra-layer model parallelism with tensor sharding along any
of its dimensions and input data partitioning across the devices. It has been proven effective in training various
state-of-the-art models. For example, GShard \cite{gshard} uses model parallelism for MoE layers and data parallelism
for other layers, and Megatron \cite{megatron} designs an SPMD strategy for the Transformer layers \cite{attention}. One
of the key benefits of SPMD parallelism is that each device executes the same program, thereby enabling scaling to a
large number of devices with a constant program compilation time.

SPMD parallelism has so far been exploited on homogeneous clusters. Enabling efficient SPMD training on a set of
heterogeneous resources facilitates better utilization of available resources for substantially lowered cost of large
model learning. Three key decisions are involved in applying SPMD parallelism in heterogeneous clusters: (i) the
\textit{sharding strategy}, i.e., deciding which dimension to partition (sharding dimension) for each tensor; (ii) the
\textit{sharding ratios} across the devices, i.e., different tensor partition sizes to assign to heterogeneous devices
according to their computation and memory capacities, to optimize device utilization; and (iii) selection of the
\textit{communication methods}, which decides the implementation of each collective communication operation for each
tensor, to best cater to different tensor sizes and different interconnect bandwidths across devices. The three
decisions are co-related. For example, if the sharding ratios are relatively even among the devices, inter-device
communication pattern resembles that of homogeneous clusters, and standard collective communication generally performs
well. In contrast, if the sharding ratio differs significantly across devices, heterogeneity-aware communication are
needed.

We propose \OurSystem, an SPMD DNN training system for heterogeneous clusters, that automatically decides optimal tensor
sharding dimension/ratios and communication methods for expedited training and optimized resource utilization. We make
the following contributions in designing \OurSystem:

$\triangleright$ We design an iterative optimization process that alternatively optimizes the SPMD sharding strategy and
sharding ratios while fixing the other one. In comparison to existing methods that only optimize each aspect once
\cite{accpar}, our iterative optimization enables us to approach the global optimum while still maintaining an
acceptable optimization time (Sec.~\ref{sec:overview}).

$\triangleright$ We novelly formulate SPMD model sharding as a program synthesis problem, to construct a distributed
program on a \textit{distributed instruction set} to emulate a given tensor program implemented on a
\textit{single-device instruction set}. We analyze the single-device program to build a background theory $\mathcal{T}$
of semantic constraints, and then employ syntax-guided synthesis \cite{syntaxguidedsynthesis} with an A*-based search
algorithm to automatically synthesize a distributed program to achieve minimal training time, that is equivalent to the
single-device program under the theory $\mathcal{T}$ (Sec.~\ref{sec:program_synthesis}).

$\triangleright$ We design a linear cost model and formulate sharding ratio optimization as a linear programming
problem, and solve it optimally with off-the-shelf solvers (Sec.~\ref{sec:load_balance}).

$\triangleright$ We explore two communication optimization techniques and integrate them into the program synthesis, to
optimize communication on heterogeneous clusters jointly with SPMD sharding. The first optimization involves the
trade-off between two \texttt{All-Gather} implementations on heterogeneous clusters and the second is to automatically
apply sufficient factor broadcasting \cite{sfb} that reduces the communication volume for certain operators
(Sec.~\ref{sec:communication_optimization}).

$\triangleright$ We implement \OurSystem on PyTorch and evaluate it on a 64-GPU heterogeneous cluster on a public cloud.
The user API of \OurSystem is analogous to the built-in DDP module of PyTorch.
Experiments with four representative image classification and language models demonstrate that \OurSystem consistently
outperforms existing systems on heterogeneous clusters and show competitive performance on homogeneous clusters while
introducing only seconds of overhead in program synthesis.

\section{Background and Motivation\label{sec:background}}

\subsection{Large Neural Networks and SPMD Parallelism}

There has been a noticeable increase in size of state-of-the-art DNN models, with examples such as GPT-3 \cite{gpt3}
containing 175 billion parameters and PaLM \cite{palme} containing 540 billion parameters. These large neural networks
are often constructed using Transformer \cite{attention} layers. Many of them incorporate Mixture-of-Expert (MoE)
\cite{moe} layers as well, which contain sparsely activated experts (each input token is processed by a fixed number of
experts, regardless of the total number of experts in an MoE layer). The tensors (parameters, gradients, activation,
optimizer states, etc.) in these models can be sizable that a single tensor exceeds the memory capacity of a modern
GPU. For example, the parameter size of an MoE layer with 2048 experts is about 36GB. Thus, partitioning the model and
deploying its shards on multiple devices is indispensable for effective training of these models.

A tensor in a DNN model is a multi-dimension array of floating-point numbers. A tensor can be partitioned (aka sharded)
into smaller tensors by splitting on any of its dimensions. For example, most of the activation tensors (intermediate
results calculated from the input to the model) in mini-batch training has a ``batch size'' dimension. Partitioning this
dimension results in the so-called data parallelism. With SPMD parallelism, the same program is executed on multiple
devices, which collectively produces a result that is identical to that of a single-device program. Existing SPMD
training systems typically partition the operators in a single-device program by choosing a sharding strategy out of a
fixed set for each operator. For example, AccPar \cite{accpar} considers three sharding strategies (sharding on the
``batch size'' dimension, ``hidden feature'' dimension, and ``reduction'' dimension) for each \texttt{MatMul} operator.
Flexflow \cite{flexflow} supports 4 dimensions of parallelism by sharding operators on the sample, operator, attribute
and parameter dimensions. When producers and consumers of a tensor are sharded in an incompatible way (i.e., a tensor is
expected by the consumer to be sharded on a different dimension than what it is produced on), collective communication
operators are inserted to switch the sharding dimension of the tensors according to pre-defined rules. For instance, if
a tensor is sharded on the batch size dimension and the consumer is expecting a full-sized tensor, \texttt{All-Gather}
can be used to gather the shards of the tensor across devices to recover the same full-sized tensor on all devices.

\subsection{Collective Communication}

\begin{figure}[t]
    \vspace{-.5em}
    \centering
    \hspace*{-3.5em}
    \includegraphics[width=.8\columnwidth]{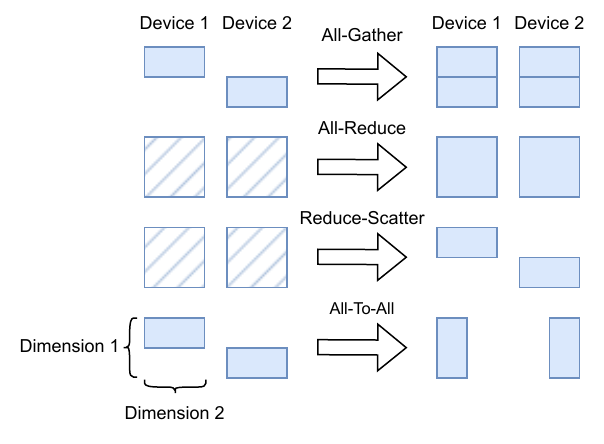}
    \vspace{-1em}
    \caption{Common collective communication operations.}\label{fig:collective}
    \vspace{-1em}
\end{figure}

Four MPI-style \cite{mpi} collective communication operations are commonly used in distributed DNN training. As
illustrated in Fig.~\ref{fig:collective}, $\texttt{All-Gather}(\bar{e}, d)$ concatenates the slices of tensor $\bar{e}$
across devices along the $d$ dimension. $\texttt{All-Reduce}(\bar{e})$ sums the replicas of a tensor $\bar{e}$ across
devices, element-wisely. $\texttt{Reduce-Scatter}(\bar{e})$ is equivalent to performing \texttt{All-Reduce} and then
sharding the results on each device, but is implemented in a more efficient way. $\texttt{All-To-All}(\bar{e},d_1,d_2)$
takes as input the tensor $\bar{e}$ that is sharded on its $d_1$ dimension and outputs the tensor that is sharded on the
$d_2$ dimension.

The current collective communication libraries have been developed for homogeneous clusters and may not exhibit optimal
performance when applied to tensors of different sizes in a cluster of different inter-device bandwidths. For instance,
NCCL \cite{nccl} requires all tensors to be of the same size for \texttt{All-Gather}. In order to perform this
communication operation on unevenly sharded tensors, the tensor shards must be first padded to the same size and
subsequently trimmed upon completion of the operation, resulting in wasted bandwidth and extra memory access.
Alternatively, \texttt{All-Gather} can be implemented with multiple \texttt{Broadcast} operations to support unevenly
sliced tensor shards without padding, at the cost of a higher kernel launching overhead.

\subsection{Syntax-Guided Synthesis}

Syntax-guided synthesis \cite{syntaxguidedsynthesis} is a type of program synthesis, where the inputs include a syntax
specification that defines the program space, a semantic correctness specification that describes the desired properties
of the synthesized program, and a background theory to verify whether a given program satisfies the semantic correctness
specification. Automated program synthesis has been successfully applied to optimize various kinds of programs, such as
SQL \cite{synthesissql} and Datalog \cite{synthesisdatalog}. Compared to the programs implemented in general-purpose
programming languages, these programs typically have simpler structures that allow easier semantic analysis. Tensor
programs that implement DNNs share the same characteristics as they are non-recursive and have no side-effects.

We exploit syntax-guided synthesis to systematically generate feasible distributed programs for SPMD parallelism, in
order to identify the best one that maximizes training speed. Most literature uses manually defined background theories,
such as the linear integer arithmetic (LIA) \cite{sygus17}. We construct a background theory for each single-device
program in the form of Hoare triples \cite{hoare} by automatically analyzing the single-device program. During the
synthesis of the distributed program, only the mathematical relations between the model output and the model inputs are
utilized. As a result, we decouple the performance of the distributed program from the implementation details of the
provided single-device program. Moreover, we automatically explore alternative implementations of operations that
achieve the same mathematical results during program synthesis.

Some existing distributed DNN training systems may be categorized as a form of "transformational program synthesis",
wherein an initial program undergoes successive modifications based on a set of rewriting rules. As an example, Unity
\cite{unity} performs pattern matching and substitution on a proposed parallel computation graph to transform programs.
Each optimized step in this process yields a complete and correct program, by the correctness of the applied rewriting
rules. In contrast, our approach diverges by exploring incomplete and potentially incorrect programs throughout the
synthesis process, and validate the semantics of the resulting distributed program using the background theory derived
from the single-device program.

\subsection{Optimal Sharding Ratios}

With pure data parallelism, the optimal sharding ratios can be readily decided by profiling the computation speed of
different devices and setting the ratios in proportion to computation speeds of devices. This does not work optimally
for SPMD parallelism where \texttt{All-Gather} and \texttt{Reduce-Scatter} are utilized to concatenate tensor shards or
aggregate tensor replicas and then shard the result across devices. In these communication operations, the devices send
and receive tensor shards of sizes proportional to the sharding ratios. The communication time therefore depends on the
size of the largest shard. Minimal communication time is achieved when the tensors are sharded evenly, as the size of
the largest shard is minimized in this case.

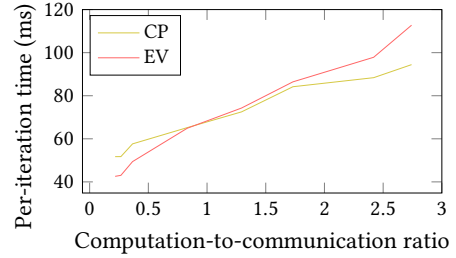
\begin{figure}[t]
    \vspace{-.2em}
    \centering
    \begin{tikzpicture}[scale=0.92, every node/.style={transform shape}]
        \begin{axis}[
            xlabel=Computation-to-communication ratio,
            ylabel=Per-iteration time (ms),
            width=.8\columnwidth,
            height=.5\columnwidth,
            xtick distance=0.5,
            xmax=3,
            ymax=120,
            legend style={at={(0.02,0.97)},anchor=north west,nodes={scale=0.85, transform shape}},
            y label style={at={(axis description cs:-0.16,1.08)},anchor=east},
        ]

        \addplot[yellow!80!black] coordinates { (0.2173099415, 51.716875) (0.266174061, 51.7378125) (0.3655061972, 57.64765625) (0.8308716987, 65.18585938) (1.293849894, 72.50140625) (1.73156235, 84.18050781) (2.419104524, 88.3825625) (2.7419397, 94.43786458) };
        \addplot[red!60] coordinates { (0.2173099415, 42.575625) (0.266174061, 43.0328125) (0.3655061972, 49.41375) (0.8308716987, 64.89070313) (1.293849894, 74.29130208) (1.73156235, 86.44425781) (2.419104524, 97.8531875) (2.7419397, 112.7417188) };

        \legend{CP, EV}
        \end{axis}
    \end{tikzpicture}
    \vspace{-.8em}
    \caption{Performance with different sharding ratios under different computation-to-communication ratios (computed for P100 GPUs).\label{fig:motivation_sharding_ratio}}
    \vspace{-1em}
\end{figure}

To demonstrate this, we train a Transformer model with intra-op model parallelism on two machines, one equipped with two
P100 GPUs and the other with two A100 GPUs. Tensors in the model are sharded across the GPUs in two ways: {\em CP}, with
sharding ratios  proportional to the computational power of the devices, and {\em EV}, evenly sharding the tensors. We
manipulate the hidden feature dimension of the model to alter the computation-to-communication ratio.
Fig.~\ref{fig:motivation_sharding_ratio} shows that when the computation time dominates, CP performs better as it
balances computation time on different devices; EV performs better when communication is the bottleneck. EV leads to
lower communication time for \texttt{All-Gather} and \texttt{Reduce-Scatter} operations and is preferable when the
computation-to-communication ratio is low. When computation time and communication time are similar, a ``sweet point''
may exist between CP and EV that achieves the optimal trade-off between load balance on different devices and fast
communication operations. Further, since different layers of a model may exhibit different computation-to-communication
ratios, the optimal sharding ratios may vary for each layer. In \OurSystem, we formulate sharding ratio optimization as
a linear programming problem to determine the optimal ratios for each part of the model.

\begin{figure*}[!t]
    \vspace{-.5em}
    \centering
    \includegraphics[width=.95\textwidth]{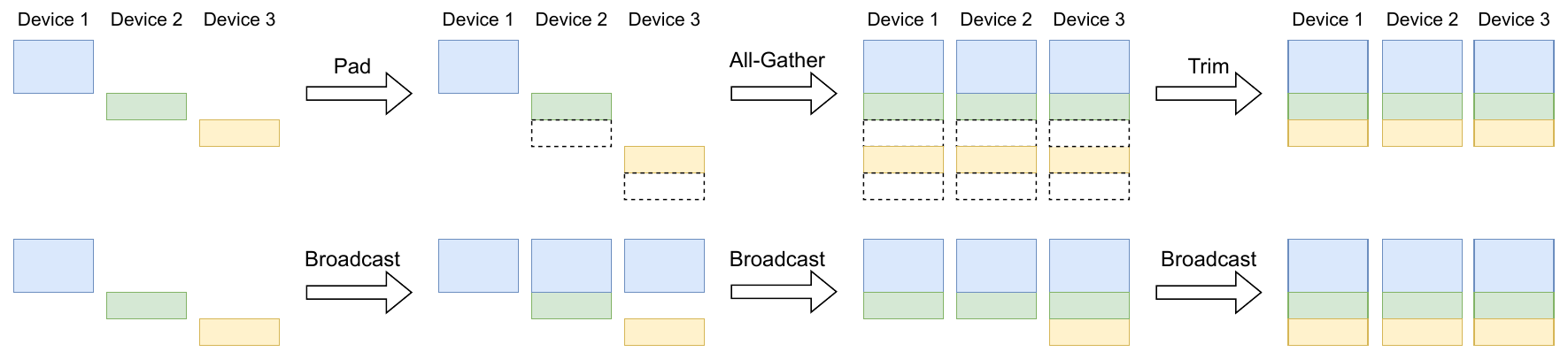}
    \vspace{-.7em}
    \caption{Different implementations of \texttt{All-Gather} for uneven shards.\label{fig:comm_opt}}
    \vspace{-.5em}
\end{figure*}

\subsection{Communication Optimization\label{sec:motivation_comm}}

When a tensor is partitioned unevenly, standard collective communication routines that assume homogeneous clusters do
not perform optimally. To explore opportunities for heterogeneity-aware communication, we study a few techniques that
can potentially benefit collective communication in heterogeneous environments and show that these optimizations need to
be jointly decided with the sharding strategy and sharding ratio selection.

\subsubsection{Padded All-Gather and Grouped Broadcast\label{sec:motivation_all_gather}}

The \texttt{All-Gather} and \texttt{Reduce-Scatter} operations implemented in NCCL require all shards to be of the same
size. To perform these operations for unevenly partitioned tensors, we can either pad the shards to the same size before
communication, or broadcast each shard seperately using a NCCL group call. Fig.~\ref{fig:comm_opt} visualizes the two
approaches.

\begin{figure}[t]
    \centering
    \begin{tikzpicture}[scale=0.92, every node/.style={transform shape}]
        \begin{axis}[
            xlabel=Maximum ratio,
            ylabel=Bandwidth (GB/s),
            width=.8\columnwidth,
            height=.5\columnwidth,
            xmin=0.24,
            xmax=1,
            xtick distance=0.1,
            ytick distance=1,
            legend style={at={(0.98,0.97)},anchor=north east,nodes={scale=0.85, transform shape}}
        ]

        \addplot[yellow!80!black] coordinates { (0.25, 6.710852845735771) (0.2575, 6.523556842305027) (0.265, 6.515296346041457) (0.2725, 6.4895474816378655) (0.28, 6.481845193809325) (0.2875, 6.23962635369891) (0.295, 6.117887778049373) (0.3025, 6.027139221869249) (0.31, 5.594152762794386) (0.3175, 5.59737506467387) (0.325, 5.465639716599261) (0.3325, 5.411601871803654) (0.33999999999999997, 5.344991931913066) (0.34750000000000003, 5.285532958309032) (0.355, 5.091989884179114) (0.3625, 5.106280368502627) (0.37, 5.084061667300449) (0.3775, 5.089533019646301) (0.385, 4.978224897318647) (0.39249999999999996, 4.853925609293298) (0.4, 4.714290192523764) (0.4075, 4.603462782849986) (0.415, 4.4995928591052765) (0.4225, 4.394302898888219) (0.43, 4.298246278735188) (0.4375, 4.268239394820867) (0.445, 4.299063306551113) (0.4525, 4.173758263633396) (0.45999999999999996, 4.165297768147031) (0.4675, 4.19941509172402) (0.475, 4.118391860797229) (0.4825, 3.9879391079382356) (0.49, 3.8949464704562313) (0.4975, 3.8125819295254986) (0.505, 3.716180192047688) (0.5125, 3.6443616740867144) (0.52, 3.672030232070043) (0.5275, 3.556837857558632) (0.5349999999999999, 3.436710141911015) (0.5425, 3.357170958164712) (0.55, 3.343881221555121) (0.5575, 3.2644862821676335) (0.565, 3.36930338757449) (0.5725, 3.3030392929420223) (0.58, 3.097554304757482) (0.5874999999999999, 3.2133767631144377) (0.595, 3.2355990263594188) (0.6025, 3.1648235764406065) (0.61, 3.057329352472138) (0.6174999999999999, 3.0977757207789685) (0.625, 3.0069114144235995) (0.6325000000000001, 2.97993309646145) (0.64, 2.97025344468875) (0.6475, 2.9560947100397175) (0.655, 2.987418675317307) (0.6625, 3.012967097894521) (0.6699999999999999, 3.014652655501361) (0.6775, 2.994660197436362) (0.685, 2.9798167799087043) (0.6925, 2.914615275753023) (0.7, 2.886938663480129) (0.7075, 2.8577674979054546) (0.715, 2.801961738236049) (0.7224999999999999, 2.7872256668731428) (0.73, 2.7858276530033454) (0.7375, 2.7274905732840375) (0.745, 2.7306555617421675) (0.7525, 2.7445699536433477) (0.76, 2.7178098122022623) (0.7675, 2.6323386764900074) (0.775, 2.6574740702602724) (0.7825, 2.597238359578336) (0.79, 2.578501576420175) (0.7975, 2.575303953434561) (0.8049999999999999, 2.606721155551143) (0.8125, 2.5943570883319276) (0.82, 2.6077928854676755) (0.8275, 2.5547045742390337) (0.835, 2.532306502520842) (0.8425, 2.531439055479927) (0.85, 2.4949693963233326) (0.8574999999999999, 2.4784225550377386) (0.865, 2.492881231673797) (0.8724999999999999, 2.467953891624501) (0.88, 2.430972189164732) (0.8875, 2.4260153937290476) (0.895, 2.338317298080992) (0.9025, 2.301947414613835) (0.9099999999999999, 2.288208699604345) (0.9175, 2.3026239193142266) (0.9249999999999999, 2.2396191192748) (0.9325, 2.3375740126392524) (0.94, 2.349126276644905) (0.9475, 2.3432980134574866) (0.955, 2.327530795407647) (0.9625, 2.3495679466352457) (0.97, 2.3140892400506363) (0.9774999999999999, 2.247074099708054) (0.985, 2.214989438107309) (0.9924999999999999, 2.037686574755153) };
        \addplot[red!60] coordinates { (0.25, 3.9814587015893945) (0.2575, 3.9562374136345415) (0.265, 3.874344947904237) (0.2725, 3.8508811103706653) (0.28, 3.8615205030963966) (0.2875, 3.9217312824375927) (0.295, 3.9031016117068984) (0.3025, 3.8184946079027715) (0.31, 3.8100175724765624) (0.3175, 3.7671573215019394) (0.325, 3.7205025624862387) (0.3325, 3.7835413488753904) (0.33999999999999997, 3.9095887389179134) (0.34750000000000003, 3.9233485726421797) (0.355, 3.9469509405603977) (0.3625, 4.022183955154728) (0.37, 3.9935402638837245) (0.3775, 3.9979187881491858) (0.385, 3.9926770023003275) (0.39249999999999996, 4.054380789615876) (0.4, 3.9792466854522206) (0.4075, 4.156509264748527) (0.415, 4.1534884818160815) (0.4225, 4.136780542850432) (0.43, 4.13582132714654) (0.4375, 4.1573956391166105) (0.445, 4.139048864368093) (0.4525, 4.155703354596898) (0.45999999999999996, 4.196792084107331) (0.4675, 4.239822124351481) (0.475, 4.313760576575417) (0.4825, 4.2713301139505315) (0.49, 4.235995410137587) (0.4975, 4.195712490541218) (0.505, 4.083607703574015) (0.5125, 3.929757211933925) (0.52, 3.952197131009126) (0.5275, 3.932704265224855) (0.5349999999999999, 3.9158395082310196) (0.5425, 3.933739955142505) (0.55, 3.952671743509118) (0.5575, 3.8492283068424467) (0.565, 3.8992866221836935) (0.5725, 3.9423876218618106) (0.58, 3.904466702815562) (0.5874999999999999, 3.8920453496863794) (0.595, 3.9107162412638723) (0.6025, 3.9335143124630556) (0.61, 3.9773441495472475) (0.6174999999999999, 3.990632831952556) (0.625, 3.9972918515490705) (0.6325000000000001, 3.9587967580657955) (0.64, 3.9114446384685473) (0.6475, 3.921452771022936) (0.655, 3.9340333218877697) (0.6625, 3.866942919830201) (0.6699999999999999, 4.00096907626095) (0.6775, 4.065652044512525) (0.685, 4.101988098432293) (0.6925, 4.169635411809526) (0.7, 4.1716627294276964) (0.7075, 4.170913044736424) (0.715, 4.255087979523359) (0.7224999999999999, 4.261840397777809) (0.73, 4.1875097338095735) (0.7375, 4.257907158979614) (0.745, 4.195233651837955) (0.7525, 4.119476650631386) (0.76, 3.977761679016341) (0.7675, 4.102971399706833) (0.775, 4.096633250393657) (0.7825, 4.201668741255197) (0.79, 4.22565093822105) (0.7975, 4.372131999711502) (0.8049999999999999, 4.383350908284565) (0.8125, 4.424602532192865) (0.82, 4.44539512653776) (0.8275, 4.4715351599816815) (0.835, 4.534897121927064) (0.8425, 4.5632600566972155) (0.85, 4.52898673536291) (0.8574999999999999, 4.495481550557043) (0.865, 4.544693725990084) (0.8724999999999999, 4.5077666539026175) (0.88, 4.4577686030997254) (0.8875, 4.507909597641691) (0.895, 4.573352039565136) (0.9025, 4.542029191359339) (0.9099999999999999, 4.626637646451007) (0.9175, 4.6507777010487565) (0.9249999999999999, 4.657302208069916) (0.9325, 4.667682739207771) (0.94, 4.7620624898076045) (0.9475, 4.684796257020891) (0.955, 4.695073408147427) (0.9625, 4.688503839342951) (0.97, 4.693569845668429) (0.9774999999999999, 4.577554467364834) (0.985, 4.676755136956468) (0.9924999999999999, 4.511262482479992) };

        \legend{Padded All-Gather, Grouped Broadcast}
        \end{axis}
    \end{tikzpicture}
    \vspace{-0.8em}
    \caption{Performance of padded \texttt{All-Gather} and grouped \texttt{Broadcast} under different tensor sharding ratios.\label{fig:comm_all_gather}}
    \vspace{-0.2em}
\end{figure}
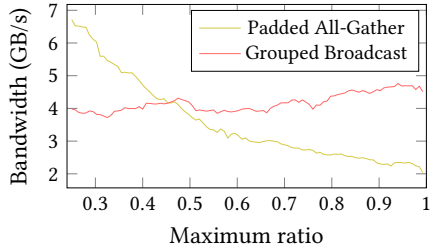

Selection between the two approaches is highly contingent upon the tensor sharding ratios employed. When the tensors are
nearly evenly partitioned, the required padding is minimal, and the padded \texttt{All-Gather} method outperforms other
approaches owing to optimizations in NCCL. In contrast, when a tensor is sharded using heavily skewed ratios among
devices, the grouped \texttt{Broadcast} approach yields better performance. Fig.~\ref{fig:comm_all_gather} illustrates
this phenomenon as we test the two approaches on a 4MB tensor in a cluster of two machines, each equipped with two
NVIDIA A100 GPUs. We allocate the largest shard to the first device and evenly partition the remaining among the other
devices. The sharding ratio on the first device then decides the skewness of sharding, depicted as the x-axis in
Fig.~\ref{fig:comm_all_gather}. The bandwidth is computed by dividing the full tensor size by the communication time,
without taking into account any padding.

As performance of the two implementations depends on the sharding ratios, when we optimize the sharding ratios, the
communication methods should be updated accordingly to achieve the optimal overall performance. In \OurSystem, we
include the selection of the two methods into the program synthesis process and interleave its optimization with the
sharding ratio optimization.

\subsubsection{Sufficient Factor Broadcasting (SFB)\label{sec:motivation_sfb}}

\begin{figure}[t]
    \centering
    \includegraphics[width=\columnwidth]{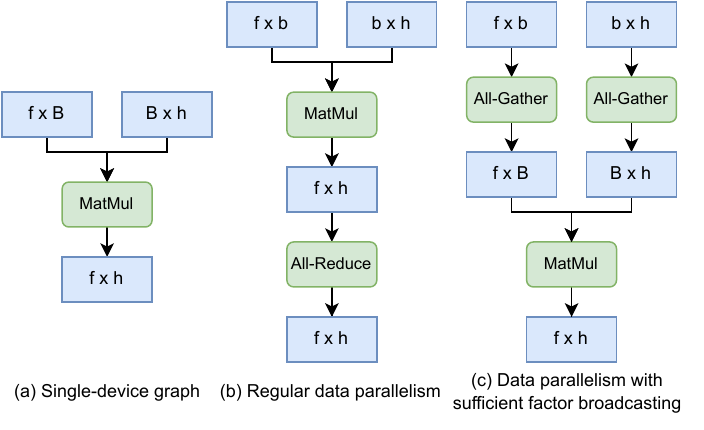}
    \vspace{-1.5em}
    \caption{Sufficient Factor Broadcasting. (b) and (c) depict the SPMD program on each device.\label{fig:sfb}}
    \vspace{-.7em}
\end{figure}

SFB \cite{sfb} exploits low-rank structures in gradient tensors to reduce parameter synchronization communication
by replacing the \texttt{All-Reduce} operation on gradient tensors with \texttt{All-Gather} operations on smaller
tensors called sufficient factors, which are sufficient to calculate the gradients. Fig.~\ref{fig:sfb} gives an example
of SFB for an \texttt{MatMul} operation. The output
is an $f\times{}h$ tensor, which is the gradient of the parameter in a fully-connected layer. The inputs are the
activation tensor of shape $b\times{}h$ and the gradient of the output with shape $f\times{}b$, where $b$ is the local
batch size on each device and $B$ is the global batch size. When $b$ is small, the gradient (the $f\times{}h$ tensor)
is not full rank, and the two aggregated tensors (the $B\times{}h$ and $f\times{}B$ tensors) are its sufficient factors
as the gradient can be calculated from the two tensors without further communication.
With standard data parallelism, the gradients are aggregated among devices with \texttt{All-Reduce}, as shown in
Fig.~\ref{fig:sfb}(b). With SFB, the input tensors are first collected with \texttt{All-Gather} and then each
device calculates the complete gradient independently. SFB changes the communication process from transferring
the gradient to transferring the sufficient factors, which can be of smaller sizes when the global batch size $B$ is
small.

The performance of SFB is primarily determined by the batch size and the number of devices involved
\cite{chilimbi,poseidon}. TAG \cite{tag} proposes an integer linear programming-based technique to automatically
identify beneficial application of SFB to tensors in a DNN model trained in a homogeneous cluster. However, uneven
tensor partitioning across heterogeneous resources introduces additional complication to this problem. As analyzed in
Sec.~\ref{sec:motivation_all_gather}, performance of \texttt{All-Gather} is influenced by the sharding ratios, on which
whether SFB is beneficial depends as well. Moreover, in the case of SFB, every device performs the \texttt{MatMul}
operation with a full batch size $B$, which may pose substantial computation overhead on slower devices. Therefore, SFB
presents a different trade-off in heterogeneous clusters. In \OurSystem, we integrate SFB into program synthesis to
ensure optimal application of SFB as we update the sharding ratios.

\section{Design Overview\label{sec:overview}}

We propose \OurSystem that jointly decides the sharding strategy, sharding ratios, and communication methods of all
tensors in a DNN model for effective SPMD model training on a heterogeneous cluster. The input to \OurSystem consists of
a single-device DNN model (DNN training program written for a single device), represented as a computation graph $(V,
E)$, and a cluster specification comprising $m$ virtual devices. A virtual device can refer to a solitary computation
unit (such as a GPU) or a small homogeneous group of physical devices (such as a machine containing multiple GPUs). We
consider the distributed training strategy at the virtual device level. In the latter case, we assume that data
parallelism is employed within each virtual device, as inter-connections within a machine typically exhibits high
bandwidth (e.g., NVLink) and data parallelism tends to yield reasonable performance.

\subsection{Main Components}
\OurSystem comprises two pivotal components: a program synthesizer and a load balancer. The program synthesizer
generates the optimal distributed program $Q$ for given sharding ratios $B$ of the tensors in the DNN model, while the load balancer produces the
optimal sharding ratios $B$ for a fixed distributed program $Q$. The distributed program $Q$ is a program on a
distributed instruction set that can be executed on all devices for distributed DNN training with the SPMD parallelism. The tensor sharding strategies and
communication methods are implicitly embedded in $Q$.

The goal is to find the optimal combination of distributed program and sharding ratios $(Q^*, B^*)$ that minimizes the
DNN per-iteration training time $t(Q, B)$. \OurSystem adopts an iterative optimization approach. During each step $s$,
we fix one of the two decision aspects and identify the optimal solution for the other:

\vspace{-4mm}
\begin{align}
    Q^{(s)} &= \operatorname*{arg\,min}_{Q}~ t(Q, B^{(s-1)}) \label{eq:overview_1}\\[-1mm]
    B^{(s)} &= \operatorname*{arg\,min}_{B}~ t(Q^{(s)}, B) \label{eq:overview_2}
\end{align}
\vspace{-3mm}

Starting with the initial sharding ratios $B^{(0)}$, which are selected to be proportional to the computation power of
devices, we first execute the program synthesizer to generate $Q^{(1)}$ following Eqn.~(\ref{eq:overview_1}). Then we
compute $B^{(1)}$ based on $Q^{(1)}$ utilizing Eqn.~(\ref{eq:overview_2}). We proceed to calculate $Q^{(2)}$ based on
$B^{(1)}$, and so on, until convergence or oscillation of the solutions is attained. In the case of oscillation, we
use the pair of $Q$ and $B$ achieving the lowest cost within the optimization process.

\begin{figure}[t]
    \vspace{-.3em}
    \centering
    \includegraphics[width=0.92\columnwidth]{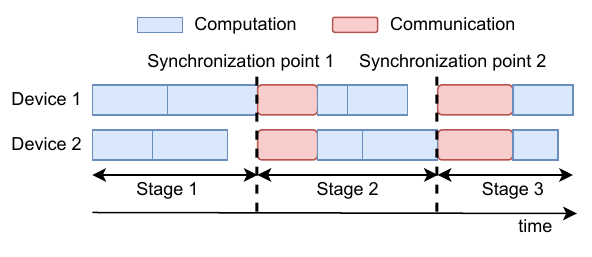}
    \vspace{-1.5em}
    \caption{Stages and synchronization points.\label{fig:cost_model_sync_point}}
    \vspace{-1em}
\end{figure}

\subsection{Cost Modeling\label{sec:cost_model}}
Directly profiling the training performance for each combination of $(Q, B)$ would be resource expensive.
We
provide an estimate of the per-iteration training time $t(Q, B)$
by simulating the execution of the distributed program $Q$ with $B$ on the heterogeneous cluster.

Collective communication typically requires every participant device to both send and receive data to and from all other participants.
It is hence reasonable to assume that all devices are synchronized before communication operations commence. This
allows us to divide the execution of a distributed program into \textit{stages}, with each stage starting with a communication
operation followed by a series of computation operations (except for the first stage which only contains computation), as illustrated in Fig.~\ref{fig:cost_model_sync_point}.
For example, the first two instructions in the program \textcircled{7} in Fig.~\ref{fig:a*_example} are in the first
stage of this program and the other two instructions are in the second stage.
All devices are synchronized at the beginning of a stage.

Let $\text{comm}^{(i)}$ and $\text{comp}_j^{(i)}$ denote the communication time and computation time of the $i$-th
stage on the $j$-th device, respectively. As each stage is globally synchronized, the iteration time is the sum of
execution time of all stages. The execution time of a stage is determined by the maximum running time of that stage on
all devices. Therefore, we have

\vspace{-1em}
$$
t(Q, B) = \sum_{i \in \text{stages}(Q)} (\text{comm}^{(i)}(B) + \max_{j \in [m]} \text{comp}_j^{(i)}(B_j))
$$
\vspace{-.6em}

\noindent where $[m] = \{1, \dots, m\}$ is the list of devices. $\text{comp}_j^{(i)}(B_j)$ can be calculated
based on the profiled flops-per-second of the $j$-th device and the estimated number of flops of the
computation operations in stage $i$.
Specifically, common operations in DNNs have numbers of flops that are linear to some of the dimensions of the
input tensors. If one of these dimensions are sharded, the number of flops of this operation on a device is proportional
to the sharding ratio of this device; otherwise, the number of flops does not change. The computation time $\text{comp}_j^{(i)}(B_j)$, which is calculated
for each operator in the $i$-th stage by dividing the flops with the flops-per-second of the $j$-th device, is therefore
a linear function of the sharding ratio $B_j$.
In the case of running \OurSystem on virtual devices which represent machines (each may contain multiple GPUs), we add
the internal communication time estimated by the internal bandwidth and parameter sizes in the stage into $\text{comp}_j^{(i)}(B_j)$.
$\text{comm}^{(i)}(B)$ is determined based on the collective operation type,
the sharding ratio $B$, and NCCL's profiling data on the cluster's network.
We run each collective operation on the cluster with tensors of different sizes and fit the latency and bandwidth
in a linear model. $\text{comm}^{(i)}(B)$ is then estimated using the fitted model, with the
input
of the tensor size of the largest shard.

\section{Distributed Program Synthesis\label{sec:program_synthesis}}

\begin{figure}[t]
    \vspace{-.3em}
    \centering
    \includegraphics[width=.98\columnwidth]{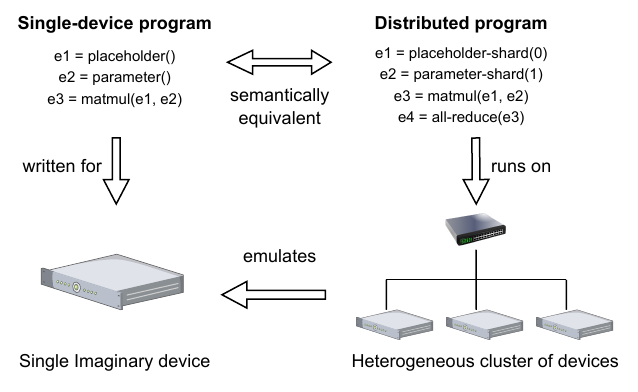}
    \vspace{-1.1em}
    \caption{A heterogeneous cluster runs distributed programs to emulate a single-device program.\label{fig:emulator}}
    \vspace{-.9em}
\end{figure}

Various optimizations such as SFB (Sec.~\ref{sec:motivation_sfb}) and different implementations of collective communications (Sec.~\ref{sec:motivation_all_gather}) can be considered in tensor sharding strategy search,
which has not been comprehensively investigated in previous SPMD training frameworks \cite{flexflow,hypar,tofu,alpa}.
In \OurSystem, we systematically approach tensor sharding strategy design in a novel way by formulating it as a program synthesis problem. Instead of
selecting a partitioning method for each operator to modify the single-device program, we synthesize a distributed
program from scratch on a distributed instruction set that emulates the single-device program, as illustrated in
Fig.~\ref{fig:emulator}. To ensure that the synthesized program is equivalent to the original single-device program, we
formalize the semantics of the single-device program and then
generate the distributed program under the constraint that the generated program must
produce a semantically equivalent output as the single-device program for any inputs.

\subsection{Distributed Programs}

A \textit{distributed program} $Q$ is defined as a sequence of symbols that follows the syntax in Fig.~\ref{fig:syntax}. An
\textit{instruction} is a computation operation or a collective communication operation with a set of tensors as inputs and produces
a tensor as the output. The computation instructions in the distributed instruction set are largely similar to the
single-device instruction set, i.e., the tensor operators provided by DNN frameworks like PyTorch \cite{pytorch}, except
for some specialized operations like \texttt{Placeholder-Shard}, which is akin to the \texttt{Placeholder} operation
utilized in single-device programs to read model inputs, but assumes that the input tensor is partitioned along a
specific dimension. Executing a distributed instruction involves executing the same instruction on all devices, where
the inputs are local tensors on each device.

\begin{figure}[t]
    \centering
    \renewcommand{\arraystretch}{1.2}
    \begin{tabular}{l c l}
        \hline
        \textit{program} & $\in$ & [\textit{instruction}] \\
        \textit{instruction} & $:=$ & \textit{computation} $\mid$ \textit{communication} \\
        \textit{computation} & $:=$ & \textit{tensor} $\gets$ \textit{optype}([\textit{tensor}]) \\
        \textit{communication} & $:=$ & \textit{tensor} $\gets$ \textit{collective}(\textit{tensor}, \textit{dim}) \\
        \textit{dim} & $\in$ & \{0, 1, \dots\} \\
        \textit{collective} & $:=$ & {\small\texttt{All-Reduce}} $\mid$ {\small\texttt{All-Gather}} $\mid$ $\dots$ \\
        \textit{optype} & $:=$ & {\small\texttt{MatMul}} $\mid$ {\small\texttt{Sigmoid}} $\mid$ $\dots$ \\
        \hline
    \end{tabular}
    \vspace{-.2em}
    \caption{Syntax of a distributed program.\label{fig:syntax}}
    \vspace{-1em}
\end{figure}

\subsection{Program Semantics\label{sec:program_semantics}}

To produce
distributed programs that are equivalent to the single-device program, we first analyze the semantics of
the single-device program to form a background theory $\mathcal{T}$, which is utilized to express the semantic
constraints during program synthesis.

The semantics of a program are expressed as a set of \textit{properties}. To formally define these properties, we
introduce {\em distributed tensors}, which are tensors produced by and used in distributed programs. A distributed tensor is a
collection of instances (i.e., shards of sharded tensors and replicas of replicated tensors) of the same tensor on all devices. A property of a distributed tensor describes its mathematical
relationship with a reference tensor, referring to a tensor in the single-device graph $(V, E)$. The properties of a distributed
tensor $\bar{e}$ are expressed as $e\mid I$, where $e \in E$ is a reference tensor and $I$ is an instruction such that
executing $I$ with $\bar{e}$ as input produces a distributed tensor whose instances on all devices are equal to $e$. For
example, if a distributed tensor $\bar{e}$ has the property $e\mid\texttt{All-Gather}(0)$, then after executing
$\texttt{All-Gather}(\bar{e}, 0)$ (where 0 denotes the sharding dimension), all devices will have a tensor that is equivalent to $e$. The set of tensors generated by
a program $Q$ is denoted by $E(Q)$, and the properties of a program $Q$ are defined as the properties of all tensors in
$E(Q)$, denoted as $P(Q)$.

The background theory $\mathcal{T}$ is expressed as a set of Hoare triples \cite{hoare}. A Hoare triple is represented
as:
$$
\hoare{precondition}{instruction}{postcondition}
$$

\noindent When the precondition is satisfied, executing the
instruction establishes the postcondition. The instruction is either a computation operation or a collective communication operation
that runs simultaneously across all devices. The precondition and postcondition are expressed as sets of properties. If
a program contains all properties in the precondition, appending the instruction to the program results in a new program
that contains the properties in the postcondition.

We derive the background theory $\mathcal{T}$ by analyzing the single-device computation graph with a set of pre-defined rules that
encodes mathematical characteristics of common tensor operations. Fig.~\ref{fig:semantics} provides some examples of
such rules on four collective communication operations and the \texttt{MatMul} operation. For conciseness, we do not explicitly name the tensors produced by an operation, but
just use $\bar{e}$ to refer to the distributed tensor that has the
property in the precondition associated with a reference tensor $e$. For example, in the first rule in
Fig.~\ref{fig:semantics}, the precondition $e\mid\texttt{All-Reduce}$ means that there is a tensor $\bar{e}$ in the program running \texttt{All-Reduce} with which produces a distributed tensor that equals a reference tensor $e$. When this condition is met, appending the instruction $\texttt{All-Reduce}(\bar{e})$ to the program leads to a new program
that meets the postcondition $e\mid\texttt{Identity}$, which means that a distributed tensor produced by the new program is
equivalent to the reference tensor $e$ (\texttt{Identity} is an operator that returns its input). As another example, the last rule in Fig.~\ref{fig:semantics} describes what is usually called reduction parallelism for \texttt{MatMul}: if $e_1$ is
sharded on its second dimension and $e_2$ is sharded on its first dimension, the \texttt{MatMul} operation can be
executed with instances of the two tensors on all devices, but an extra \texttt{All-Reduce} needs to be used on the
result to obtain a tensor that equals the single-device \texttt{MatMul}.

The background theory $\mathcal{T}$ is obtained for the single-device graph $(V, E)$ by enumerating all rules and
finding matches in the single-device graph, and then gathering the Hoare triples from the matched rules.
The semantic constraint of the distributed program $Q$ is defined as $(l\mid\texttt{All-Reduce}) \in P(Q)$,
where $l$ is the output tensor (typically the training loss) of the single-device graph.
If a distributed program can be proved to have property $l\mid\texttt{All-Reduce}$ under theory $\mathcal{T}$, it is deemed equivalent to the single-device graph as it produces
the same output as the single-device graph.
We will use such semantic constraints to produce equivalent distributed programs with different tensor sharding and communication strategies.

\begin{figure}[t]
    \vspace{-.3em}
    \begin{scprooftree}{0.8}
        \AxiomC{\(\forall e \in E\)}
        \UnaryInfC{\(\hoare{e\mid\texttt{All-Reduce}}{\texttt{All-Reduce}(\bar{e})}{e\mid\texttt{Identity}}\)}
    \end{scprooftree}

    \vspace{-.3em}
    \begin{scprooftree}{0.8}
        \AxiomC{\(\forall e \in E,~ \forall d \in \operatorname{dims}(e)\)}
        \UnaryInfC{\(\hoare{e\mid\texttt{All-Reduce}}{\texttt{Reduce-Scatter}(\bar{e}, d)}{e\mid\texttt{All-Gather}(d)}\)}
    \end{scprooftree}

    \vspace{-.3em}
    \begin{scprooftree}{0.8}
        \AxiomC{\(\forall e \in E,~ \forall d_1, d_2 \in \operatorname{dims}(e),~ d_1 \neq d_2\)}
        \UnaryInfC{\(\hoare{e\mid\texttt{All-Gather}(d_1)}{\texttt{All-To-All}(\bar{e}, d_1, d_2)}{e\mid\texttt{All-Gather}(d_2)}\)}
    \end{scprooftree}

    \vspace{-.3em}
    \begin{scprooftree}{0.8}
        \AxiomC{\(\forall e \in E,~ \forall d \in \operatorname{dims}(e)\)}
        \UnaryInfC{\(\hoare{e\mid\texttt{All-Gather}(d)}{\texttt{All-Gather}(\bar{e}, d)}{e\mid\texttt{Identity}}\)}
    \end{scprooftree}

    \vspace{-.7em}
    \begin{scprooftree}{0.8}
        \AxiomC{\(\forall e_1,e_2,e_3 \in E,~ e_3 = \texttt{MatMul}(e_1, e_2)\)}
        \UnaryInfC{\(\hoare{e_1\mid\texttt{All-Gather}(0),~ e_2\mid\texttt{Identity}}{\texttt{MatMul}(\bar{e}_1,\bar{e}_2)}{e_3\mid\texttt{All-Gather}(0)}\)}
    \end{scprooftree}

    \vspace{-.6em}
    \begin{scprooftree}{0.8}
        \AxiomC{\(\forall e_1,e_2,e_3 \in E,~ e_3 = \texttt{MatMul}(e_1, e_2)\)}
        \UnaryInfC{\(\hoare{e_1\mid\texttt{Identity},~ e_2\mid\texttt{All-Gather}(1)}{\texttt{MatMul}(\bar{e}_1,\bar{e}_2)}{e_3\mid\texttt{All-Gather}(1)}\)}
    \end{scprooftree}

    \vspace{-.6em}
    \begin{scprooftree}{0.78}
        \AxiomC{\(\forall e_1,e_2,e_3 \in E,~ e_3 = \texttt{MatMul}(e_1, e_2)\)}
        \UnaryInfC{\(\hoare{e_1\mid\texttt{All-Gather}(1),~ e_2\mid\texttt{All-Gather}(0)}{\texttt{MatMul}(\bar{e}_1,\bar{e}_2)}{e_3\mid\texttt{All-Reduce}}\)}
    \end{scprooftree}
    \vspace{-.5em}
    \caption{Examples of the semantics of common collective communication operations and the \texttt{MatMul} operation.\label{fig:semantics}}
    \vspace{-1em}
\end{figure}

\subsection{Program Search Algorithm\label{sec:search_algorithm}}

A naive way to generate the best distributed program $Q$ (that minimizes the iteration time with $B$) is to enumerate all possible programs, produced by sharding tensors along different dimensions and using different suitable collective communication following the syntax in
Fig.~\ref{fig:syntax}. We can produce the programs in a breadth-first search manner and verify if a result program is semantically equivalent to the
single-device program.
However, the number of possible distributed programs grows exponential with the number of instructions. The exhaustive search is impractical for DNN models with hundreds or more operators.

We propose a more efficient program search algorithm based on the following ideas:
(i) we estimate a cost lower-bound (execution time) for a partial program and stop searching further based on this partial program if its cost lower-bound is higher than the current best program;
(ii) if two programs lead to the same set of properties, we discard the one with a higher cost.
A program $Q$ is considered \textit{complete} if $l\mid\texttt{All-Reduce} \in P(Q)$, indicating that it is already
semantically equivalent to the single-device program and no additional instructions are required. Programs constructed
during the search that are not complete are called partial programs.

We use A* algorithm combined with the idea of dynamic programming to search for the optimal distributed program $Q^*$,
as given in Fig.~\ref{fig:a*}.
We maintain a priority queue $S$ that contains partial programs and their
scores. The score of a partial program $Q$ is an estimate of the per-iteration execution time of the optimal complete program $Q_c$ that
starts with $Q$: $\text{score}(Q) = \text{cost}(Q) + \text{ecost}(Q)$, where $\text{cost}(Q)$ is the execution time of the partial program $Q$ and $\text{ecost}(Q)$ is a heuristic function that
estimates the future cost of the program $\text{cost}(Q_c) - \text{cost}(Q)$. In order for the A* algorithm to find the optimal
program, the heuristic function ecost must not overestimate the future cost (i.e., we need $\text{ecost}(Q) \leq
\text{cost}(Q_c) - \text{cost}(Q)$), as otherwise the program will be excluded from the potential solutions.
We use the minimum required execution time of program $Q$ as its ecost, assuming infinite communication
bandwidth among the devices.
For a complete program $Q$, $\text{cost}(Q) = t(Q, B)$, obtained via our cost modeling in Sec.~\ref{sec:cost_model}. For an incomplete program, $\text{cost}(Q)$ is calculated
similarly, but only includes the time to reach the last synchronization point (e.g., synchronization point 2 in
Fig.~\ref{fig:cost_model_sync_point}).

In each loop of the algorithm, we retrieve the program $Q$ with the lowest score from the priority queue $S$ (Line 6) and find an
instruction that can be appended to it (Line 7). We want the resulting program to have more properties than $Q$; otherwise, the resulting program would be strictly worse than $Q$ because it contains more instructions (therefore a higher
cost) but is not closer to a complete program. Therefore, we enumerate the Hoare triples in $\mathcal{T}$ and find instructions whose
precondition is met but postcondition contains properties not
in $P(Q)$ (Line 7), so appending the
instructions to $Q$ is guaranteed to produce a new program $Q'$ that contains more properties than $Q$. Next, we check
if there are programs that are strictly better than $Q'$ and stops further constructing programs based on  $Q'$ if so (Lines 9 to 11). We also
remove any programs in $S$ that are strictly worse than $Q'$
(Lines 12 to 14).
Finally, if $Q'$ is complete, we compare it with the current best complete program and replace the best program if $Q'$
is better (Line 16). If $Q'$ is not complete, we add it to $S$ and proceed to next loop (Line 18).

\begin{figure}[t]
\centering
\begin{minipage}{\linewidth}
    \small
    \begin{algorithmic}[1]
        \STATE \textbf{Input: } computation graph $(V, E)$, sharding ratios $B$
        \STATE \textbf{Output: } Optimal distributed
        program $Q^*$
        \item[]
        \STATE Initialize a priority queue $S$ with an empty program $Q_\emptyset$
        \STATE Initialize best program $Q^* = \textbf{null}$ and set $\text{cost}(Q^*) = \infty$
        \WHILE {$\exists Q \in S, \text{score}(Q) < \text{cost}(Q^*)$}
            \STATE Remove the program $Q$ with lowest score from $S$
            \FOR{$\hoare{precondition}{instruction}{postcondition} \in \mathcal{T}$ of the single-device program where $precondition \subseteq P(Q)$ and $postcondition \nsubseteq P(Q)$}
                \STATE $Q' = Q \cup instruction$;\quad $P(Q') = P(Q) \cup postcondition$
                \IF {$\exists Q_s \in S$ s.t. $P(Q_s) \supseteq P(Q')$ and $\text{cost}(Q_s) \leq \text{cost}(Q')$}
                    \STATE \textbf{continue}
                \ENDIF
                \FOR{$Q_s \in S$ where $P(Q') \supseteq P(Q_s)$ and $\text{cost}(Q') \leq \text{cost}(Q_s)$}
                    \STATE remove $Q_s$ from $S$
                \ENDFOR
                \IF {$Q'$ is \textit{complete}}
                    \STATE $Q^* \leftarrow Q'$ if $\text{cost}(Q') < \text{cost}(Q^*)$
                \ELSE
                    \STATE add $Q'$ into $S$
                \ENDIF
            \ENDFOR
        \ENDWHILE
    \end{algorithmic}
\end{minipage}
\vspace{-.2em}
\caption{A* algorithm for sharding strategy search\label{fig:a*}}
\vspace{-1em}
\end{figure}

The resulting program contains both computation operations and communication operations (as in instructions). The
sharding strategy is implicitly included in the generated program when synthesizing communication operations and special
operations like \texttt{Placeholder-Shard}. Following our semantic constraints, the generated collective communication
is guaranteed to properly handle the tensors the computation produces. For example, $\texttt{All-Gather}(\bar{e_1}, 0)$
will only be generated upon a tensor $\bar{e_1}$ that is previously sharded on its $0$-th dimension. Similarly,
$\texttt{All-Reduce}(\bar{e_2})$ will only be generated if performing this operation produces a tensor that equals a
tensor $e_2$ in the single-device graph.

\begin{figure*}[t]
    \vspace{-.5em}
    \centering
    \includegraphics[width=.99\textwidth]{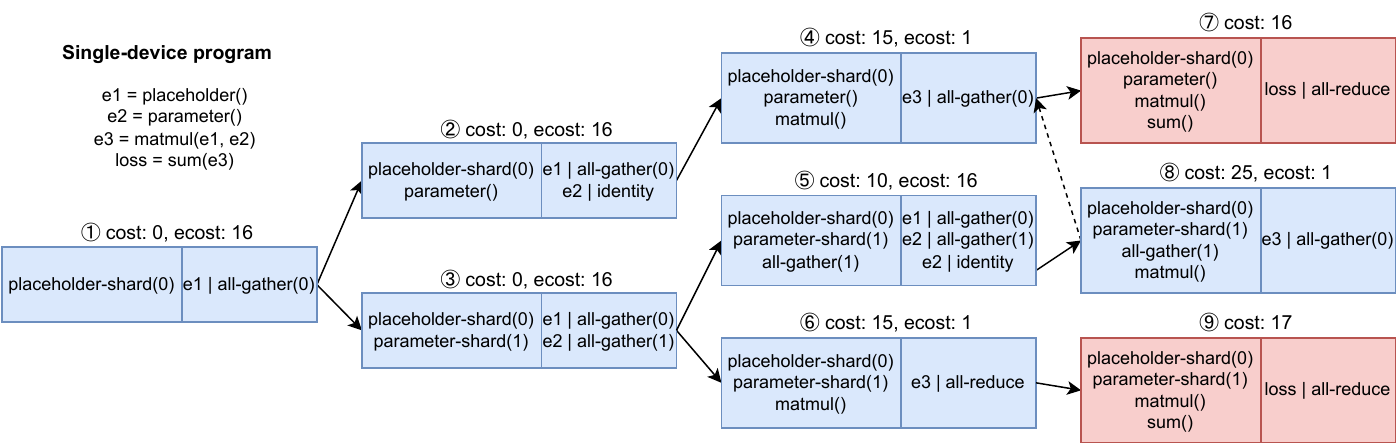}
    \vspace{-.5em}
    \caption{A* search example. Names of distributed tensors (e.g., $\bar{e}_1$) are omitted.\label{fig:a*_example}}
    \vspace{-.5em}
\end{figure*}

We give an example of the searching process in Fig.~\ref{fig:a*_example}. Suppose that the single-device program
contains 4 instructions, as given on the top left in the figure. The \texttt{Placeholder} operation retrieves a batch of
input samples. The \texttt{Parameter} operation loads a parameter tensor of the model. Then the matrix product of the
two tensors are computed and its element sum is computed as the loss. Starting with an empty distributed program, assume
that we find the following matching rule:

\vspace{-.3em}
\begin{scprooftree}{0.85}
    \AxiomC{\(\forall e \in E,~ e = \texttt{Placeholder}()\)}
    \UnaryInfC{\(\hoare{\emptyset}{\texttt{Placeholder-Shard}(0)}{e\mid\texttt{All-Gather}(0)}\)}
\end{scprooftree}

\noindent whose precondition is met (as it has no properties). We append the instruction $\texttt{Placeholder-Shard}(0)$
to the empty program and obtain program \textcircled{1}. Suppose there is no other rule with an empty precondition;
\textcircled{1} is now the only element in the priority queue. For brevity, we only consider partitioning $e_1$ on the
first dimension (dimension 0) and $e_2$ on its second dimension (dimension 1), and do not include communication
optimizations in this example. In the second loop, we retrieve \textcircled{1} from the priority queue and append
different instructions to it by enumerating $\mathcal{T}$, leading to programs \textcircled{2} and \textcircled{3}. In
the third loop, we retrieve \textcircled{2} from the priority queue and append a \texttt{MatMul} operation and obtain
program \textcircled{4}. Note that we remove the properties regarding $e_1$ and $e_2$ as the two tensors will no longer
be used in the rest of the program. We will introduce details of this optimization in
Sec.~\ref{sec:search_algorithm_optimizations}. Then in the fourth loop, we find \textcircled{5} and \textcircled{6}
based on \textcircled{3}. In the fifth loop, we obtain program \textcircled{7} which is a complete program as it has
property $loss\mid\texttt{All-Reduce}$. Since its cost is no higher than the scores of \textcircled{5} and
\textcircled{6} (the two programs in the priority queue), the search terminates and returns \textcircled{7} as the
optimal program.

\subsection{Communication Optimization\label{sec:communication_optimization}}

With our program synthesis approach, we can readily incorporate the two communication optimization techniques
(Sec.~\ref{sec:motivation_comm}) into the distributed program search, to jointly optimize communication on heterogeneous
clusters with the sharding strategy.

As discussed in Sec.~\ref{sec:motivation_all_gather}, there can be two implementations of \texttt{All-Gather} on a
heterogeneous cluster, which exhibit different performance under different sharding ratios. To automatically decide the
better implementation under a given sharding ratio $B$, we can add the following rule during program search:

\vspace{-.3em}
\begin{scprooftree}{0.85}
    \AxiomC{\(\forall e \in E,~ \forall d \in \operatorname{dims}(e)\)}
    \UnaryInfC{\(\hoare{e\mid\texttt{All-Gather}(d)}{\texttt{Grouped-Broadcast}(\bar{e}, d)}{e\mid\texttt{Identity}}\)}
\end{scprooftree}

\noindent The rule has the same precondition and postcondition as the \texttt{All-Gather} instruction in
Fig.~\ref{fig:semantics} (which refers to the padded \texttt{All-Gather} implementation), but indicates using multiple
Broadcast operations to implement \texttt{All-Gather}. Whenever a partial program meets the precondition, our A* search
will attempt both instructions and retain only the one with better estimated performance, using lines 9 to 14 in
Fig.~\ref{fig:a*}.

To support sufficient factor broadcasting (Fig.~\ref{fig:sfb}(c)), we only need to add the following rule in additional
to those in Fig.~\ref{fig:semantics}:

\vspace{-.3em}
\begin{scprooftree}{0.85}
    \AxiomC{\(\forall e_1,e_2,e_3 \in E,~ e_3 = \texttt{MatMul}(e_1, e_2)\)}
    \UnaryInfC{\(\hoare{e_1\mid\texttt{Identity},~ e_2\mid\texttt{Identity}}{\texttt{MatMul}(\bar{e}_1,\bar{e}_2)}{e_3\mid\texttt{Identity}}\)}
\end{scprooftree}

\noindent which denotes that all devices duplicate the same computation with identical input data. By applying this rule
and the fifth rule in Fig.~\ref{fig:semantics} to the single-device program in Fig.~\ref{fig:sfb}(a), which is inside
the search space of our A* algorithm, we can generate the program depicted in Fig.~\ref{fig:sfb}(c). By adding similar
rules to common operators in DNNs, \OurSystem's program synthesis process can automatically explore other possible
applications of SFB.

\subsection{Search Time Optimization\label{sec:search_algorithm_optimizations}}

As the number of operations increases, the execution time of our A* algorithm may still be substantial. We further
propose three heuristics to balance the search time and performance of the obtained sharding strategy.

Our first optimization involves fusing Hoare triples that have empty preconditions with their consumers. For instance,
for a \texttt{Placeholder} operation in the single-device graph that produces reference tensor $e$, we may create a
Hoare triple $\hoare{\emptyset}{\texttt{Placeholder}}{e\mid\texttt{Identity}}$. Since it has an empty precondition, the
code may appear at any position in the program before the first consumer of $e$, and our search algorithm would explore
all possible positions of such instructions during the search. To reduce the overhead, we fuse those Hoare triples with
their consumers to generate new Hoare triples with two consecutive instructions. Specifically, for two Hoare triples
$\hoare{Pre_1}{Inst_1}{Post_1}$ and $\hoare{Pre_2}{Inst_2}{Post_2}$, if $Pre_1 = \emptyset$ and $Post_1 \subseteq
Pre_2$, we remove the first triple from $\mathcal{T}$ and insert a new Hoare triple $\hoare{Pre_2 \setminus
Post_1}{Inst_1 \cup Inst_2}{Post_1 \cup Post_2}$. This ensures that all instructions with empty preconditions occur
directly before their first consumers and eliminates the enumeration of their positions in the program.

Our second optimization is to disallow repeated communications of the same reference tensor. We also disallow
communication of tensors produced by \texttt{Placeholder} and \texttt{Parameter}, which can directly produce sharded
tensors with specialized instructions, \texttt{Placeholder-Shard} and \texttt{Parameter-Shard}. Without this
optimization, \OurSystem attempts to append multiple communication instructions for each tensor because they introduce
new properties to the program, even though most of these properties are not utilized. For a Hoare triple that generates
a communication instruction of reference tensor $e$, we append a special property $e\mid\neg \texttt{Communicated}$ to
its precondition and $e\mid\texttt{Communicated}$ to its postcondition. This makes communication instructions of the
same reference tensor conflict with each other, so that at most one of them can appear in one distributed program.

Our third optimization is about removing redundant properties from partial programs to increase the number of programs
we can prune in lines 9 to 14 in Fig.~\ref{fig:a*}. A property is redundant to a partial program $Q$ if it does not
appear in the precondition of any instruction whose $postcondition \nsubseteq P(Q)$. For example, the property $e_1 \mid
\texttt{All-Gather}(0)$ in the partial program \textcircled{2} only appears in the preconditions of two kinds of
instructions: communication operations of $e_1$ and $\texttt{MatMul}(e_1, e_2)$. The former are not considered as a
result of our second optimization. Therefore, after inserting the $\texttt{MatMul}$ instruction to form \textcircled{4},
no instruction with $e_1 \mid \texttt{All-Gather}(0)$ in its precondition satisfies $postcondition \nsubseteq
P(\text{\textcircled{4}})$, and we can safely remove this property from $P(\text{\textcircled{4}})$ without affecting
the final result.

\section{Load Balancing\label{sec:load_balance}}

We next detail our design of the load balancer that produces the optimal sharding ratios $B$
for a fixed distributed program $Q$, i.e., solve $\operatorname*{arg\,min}_{B} t(Q, B)$.

\subsection{A Base Case\label{sec:load_balance_base_case}}

We first consider a basic case where the same sharding ratios across the devices are used for each tensor in the DNN
model. As a collective communication operation involves all devices and is bottlenecked by the slowest participant, the
communication time in the $i$-th stage (Sec.~\ref{sec:cost_model}), $\text{comm}^{(i)}(B)$, is decided by the largest
communication time of a tensor shard, i.e., $\text{comm}^{(i)}(B)$ is a linear function of $\max_{j \in [m]}(B_j)$.
Since we synchronize all devices at the beginning of each stage and the collective communication operations take the
same time across devices (Fig.~\ref{fig:cost_model_sync_point}), the computation time of the $i$-th stage is the maximum
computation time among the devices, i.e., $\max_j \text{comp}_j^{(i)}(B_j)$. We then solve the following problem to
obtain $B$:

\noindent\begin{minipage}{\columnwidth}
\vskip .5em
\begin{equation}
    \min~~ \sum_{i \in \text{stages}(Q)}
    (\text{comm}^{(i)}(B) + \max_{j \in [m]} \text{comp}_j^{(i)}(B_j)) \label{eqn:opt_B}
\end{equation}
\vskip -1.2em
\begin{align*}
    \text{subject to:} &\quad \sum_{j = 1}^{m} B_j = 1, \\
                       &\quad B_j \geq 0,\quad \forall j \in [m] \\[-1mm]
\end{align*}
\end{minipage}
\vspace{-1em}

\noindent The objective function is the sum of the communication time and computation time of all stages, which is the per-iteration training
time that we are minimizing.
The constraints state that sharding ratios are non-negative and sum to 1. In \OurSystem, the functions
$\text{comm}^{(i)}$ and $\text{comp}_j^{(i)}$ are modeled as linear functions on bandwidth and flops
(Sec.~\ref{sec:cost_model}). Therefore, the optimization problem is a linear program and can be solved efficiently with
off-the-shelf solvers.

After obtaining the optimal fractional solutions, we round the sharded sizes of each tensor to integers and ensure they
add up to the total length of the dimension that is sharded on. We first set the sharded sizes to their nearest
integers. If the sum is larger or smaller than the original size, we repeatedly reduce/increase the size by one for a
shard that introduces smallest rounding errors, until the sizes of the sharded tensors sum to the original tensor.

\subsection{Different Sharding Ratios across the Model}

If the DNN model contains many layers and the computation-communication ratio differs across layers, using the same
sharding ratios throughout the model may not be ideal. Due to the large number of tensors in a model, computing a
different set of sharding ratios for each tensor may incur high computation overhead. Instead, we partition the tensors
in the model, $E$, into $g$ segments, denoted by $E_k, 1 \leq k \leq g$, and identify the sharding ratios for each model
segment. The segment division can be either specified by the user (such as using the layers of the model) or determined
using a partition algorithm such as METIS \cite{metis} (which minimizes the tensor size on the cuts while balancing the
size of partitions). The sharding ratios $B$ subsequently become a $g \times m$ matrix, where $B_{k,j}$ represents the
sharding ratio for tensors in the $k$-th model segment on the $j$-th device.

For a tensor in the $k$-th segment that is produced by tensors from other segments (whose sharding ratios may not be
$B_k$), an \texttt{All-To-All} operation is inserted to coordinate between the sharding ratios. To simplify
implementation, we always insert \texttt{All-To-All} operations at the boundaries of segments, regardless of whether the
sharding ratios are the same or not between the two segments. As a result, there are synchronization points at segment
boundaries, and each stage is entirely within a model segment. In this way, for each segment, we can solve the
optimization problem in (\ref{eqn:opt_B}) independently to determine the optimal sharding ratios for tensors in that
segment.

\section{Implementation\label{sec:implementation}}

\begin{figure}[t]
    \centering
    \includegraphics[width=0.75\columnwidth]{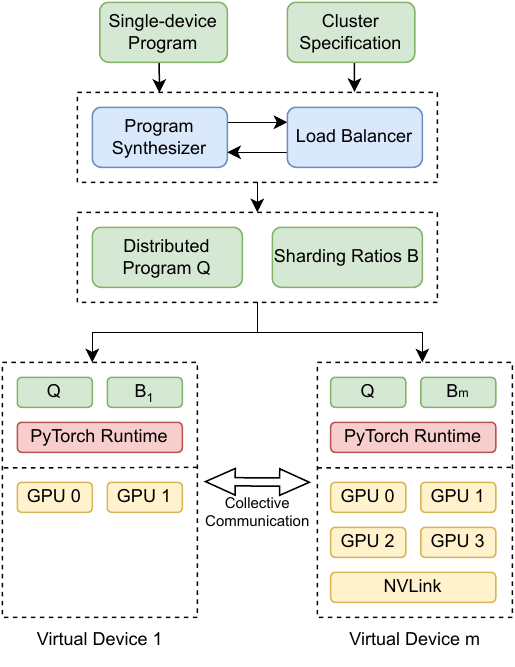}
    \vspace{-1em}
    \caption{\OurSystem Implementation.\label{fig:arch}}
    \vspace{-1em}
\end{figure}

\OurSystem is implemented on PyTorch 1.13.1 \cite{pytorch} with 2789 lines of Rust code for the program synthesizer, 69
lines of Rust code for the load balancer, and 362 lines of Python code for the profiler, collective operations, and the
user API. Fig.~\ref{fig:arch} shows the modules we implemented in \OurSystem.
The single-device program is
represented as a PyTorch fx \cite{torchfx} graph. The cluster specification contains the information of the
virtual devices (GPUs and machines), including the profiled flops-per-second of the devices and the latency and bandwidth of each collective primitives on this cluster. The program synthesizer (Sec.~\ref{sec:program_synthesis}) and the load balancer (Sec.~\ref{sec:load_balance}) are run on CPU to identify the optimal distributed program $Q$ and sharding ratio $B$.
We use CBC \cite{cbc} to solve the sharding ratio optimization problem.

At the begining of model training, \OurSystem broadcasts $Q$ and $B$ to all workers (virtual devices), which run them on the PyTorch runtime. Each worker first initializes the
original single-device model in CPU using the same seed. For each $\texttt{Parameter-Shard}(d)$ operation in $Q$, the $j$-th
worker shards the corresponding parameter along its $d$ dimension and only keeps the slice corresponding to the portion of $\sum_{x=1}^{j-1} B_x$
to $\sum_{x=1}^{j} B_x$. The sharded parameters are loaded to GPU for training.

In each training iteration, the workers each load a minibatch of input data according to their sharding ratios. Then they execute $Q$ and synchronize with each other when executing collective communications. After running $Q$, each worker applies the
gradient to its own parameter shards.
The collective communication operations are implemented using PyTorch's API with the NCCL \cite{nccl} backend.

When \OurSystem is run on a virtual device which represents a machine, program $Q$ sent to the machine is replicated to all GPUs in the
machine. Regular data-parallel training is carried out among GPUs in the machine. Each collective operation in $Q$ is replaced by a three-step
communication operation: the tensors for communication are first aggregated from all GPUs to GPU 0 using \texttt{Gather}
or \texttt{Reduce}; GPU 0 participants in the global collective communication using the aggregated tensor; and then
GPU 0 broadcasts the result to other GPUs in the machine using \texttt{Scatter} or \texttt{Broadcast}.

The user API of \OurSystem is analogous to the built-in DDP module of PyTorch: the user calls \texttt{hap.HAP} function with
a single-device PyTorch model and a Python Dict of device specification, and the function returns a distributed model
that can run on the cluster with PyTorch's \texttt{torch.distributed} module.
We plan to open-source \OurSystem to the community.

\section{Evaluation\label{sec:evaluation}}

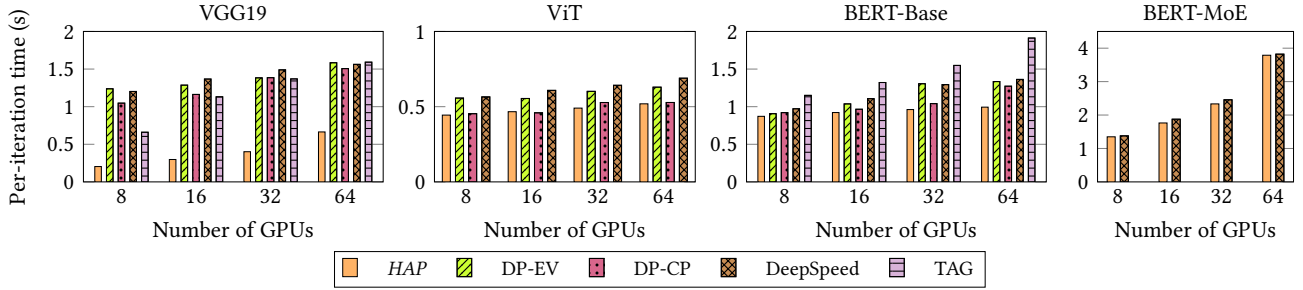
\begin{figure*}[t]
    \centering
    \begin{tikzpicture}[ scale=0.97, every node/.style={transform shape} ]
        \begin{axis}[name=vgg,
            title = {\small VGG19},
            title style={yshift=-.5em},
            ybar, ymin=0, bar width=.09cm,
            enlarge x limits=0.17,
            width=.32\textwidth,
            height=.205\textwidth,
            symbolic x coords={8, 16, 32, 64},
            xtick style={draw=none},
            xtick distance=1,
            ymax=2,
            ytick distance=0.5,
            x tick label style={yshift={0.5em}},
            xlabel={\small Number of GPUs},
            ylabel={\small Per-iteration time (s)},
            legend image code/.code={ \draw [#1] (0cm,-0.1cm) rectangle (0.2cm,0.15cm); },
            legend style={nodes={scale=0.8, transform shape}, column sep = 10pt, legend columns = -1, legend to name = legend}]

            \addplot[style={fill=orange!60}] coordinates { (8,0.2040309906) (16,0.2976762891) (32,0.4011395097) (64,0.6641462564) };
            \addplot[style={fill=lime!80,postaction={pattern=north east lines}}] coordinates { (8,1.237666988) (16,1.286270475) (32,1.383243549) (64,1.583955352) };
            \addplot[style={fill=purple!60,postaction={pattern=dots}}] coordinates { (8,1.046966016) (16,1.162634134) (32,1.385361719) (64,1.506026478) };
            \addplot[style={fill=brown!90,postaction={pattern=crosshatch}}] coordinates { (8,1.202265024) (16,1.367312551) (32,1.489451003) (64,1.562532985) };
            \addplot[style={fill=violet!30,postaction={pattern=horizontal lines}}] coordinates { (8,0.6600243574) (16,1.130984866) (32,1.369305157) (64,1.591682348) };
            \addlegendentry{\OurSystem}
            \addlegendentry{DP-EV}
            \addlegendentry{DP-CP}
            \addlegendentry{DeepSpeed}
            \addlegendentry{TAG}
        \end{axis}
        \begin{axis}[name=vit,
            at={($(vgg.north east)+(2em,0)$)}, anchor=north west,
            title = {\small ViT},
            title style={yshift=-.5em},
            ybar, ymin=0, bar width=.11cm,
            enlarge x limits=0.16,
            width=.29\textwidth,
            height=.205\textwidth,
            symbolic x coords={8, 16, 32, 64},
            xtick style={draw=none},
            xtick distance=1,
            ymax=1,
            ytick distance=0.5,
            x tick label style={yshift={0.5em}},
            xlabel={\small Number of GPUs}]

            \addplot[style={fill=orange!60}] coordinates { (8,0.4439690948) (16,0.4665999413) (32,0.4904685974) (64,0.5189594865) };
            \addplot[style={fill=lime!80,postaction={pattern=north east lines}}] coordinates { (8,0.557106483) (16,0.5545019388) (32,0.6018322706) (64,0.6296544552) };
            \addplot[style={fill=purple!60,postaction={pattern=dots}}] coordinates { (8,0.4533306003) (16,0.4591528654) (32,0.5277118802) (64,0.5278520703) };
            \addplot[style={fill=brown!90,postaction={pattern=crosshatch}}] coordinates { (8,0.564966507) (16,0.6083509636) (32,0.6420982361) (64,0.6893313766) };
        \end{axis}
        \begin{axis}[name=bert,
            at={($(vit.north east)+(2em,0)$)}, anchor=north west,
            title = {\small BERT-Base},
            title style={yshift=-.5em},
            ybar, ymin=0, bar width=.09cm,
            enlarge x limits=0.17,
            width=.32\textwidth,
            height=.205\textwidth,
            symbolic x coords={8, 16, 32, 64},
            xtick style={draw=none},
            xtick distance=1,
            ymax=2,
            ytick distance=0.5,
            x tick label style={yshift={0.5em}},
            xlabel={\small Number of GPUs}]

            \addplot[style={fill=orange!60}] coordinates { (8,0.8716604471) (16,0.9221706629) (32,0.9618194103) (64,0.9938067675) };
            \addplot[style={fill=lime!80,postaction={pattern=north east lines}}] coordinates { (8,0.9068121672) (16,1.037033498) (32,1.305151966) (64,1.333852756) };
            \addplot[style={fill=purple!60,postaction={pattern=dots}}] coordinates { (8,0.9209309101) (16,0.9658694506) (32,1.04034332) (64,1.272203076) };
            \addplot[style={fill=brown!90,postaction={pattern=crosshatch}}] coordinates { (8,0.9721413374) (16,1.106371427) (32,1.293376148) (64,1.362542236) };
            \addplot[style={fill=violet!30,postaction={pattern=horizontal lines}}] coordinates { (8,1.150283547) (16,1.321393755) (32,1.5485550057) (64,1.91239063188) };
        \end{axis}
        \begin{axis}[name=moe,
            at={($(bert.north east)+(2em,0)$)}, anchor=north west,
            title = {\small BERT-MoE},
            title style={yshift=-.5em},
            ybar, ymin=0, bar width=.11cm,
            enlarge x limits=0.13,
            width=.24\textwidth,
            height=.205\textwidth,
            symbolic x coords={8, 16, 32, 64},
            xtick style={draw=none},
            xtick distance=1,
            ymax=4.5,
            ytick distance=1,
            x tick label style={yshift={0.5em}},
            xlabel={\small Number of GPUs}]

            \addplot[style={fill=orange!60}] coordinates { (8,1.34898026) (16,1.762865591) (32,2.333233702) (64,3.788795328) };
            \addplot[style={fill=brown!90,postaction={pattern=crosshatch}}] coordinates { (8,1.377427089) (16,1.875422978) (32,2.456089568) (64,3.821276152) };
        \end{axis}
        \node at ($(7.9cm,-1.2cm)$) {\ref{legend}};
    \end{tikzpicture}

    \vspace{-.8em}
    \caption{Per-iteration training time on heterogeneous clusters.\label{fig:exp_het}}
    \vspace{-.8em}
\end{figure*}

\begin{figure*}[t]
    \centering
    \begin{tikzpicture}[ scale=0.97, every node/.style={transform shape} ]
        \begin{axis}[name=vgg,
            title = {\small VGG19},
            title style={yshift=-.5em},
            ybar, ymin=0, bar width=.095cm,
            enlarge x limits=0.17,
            width=.32\textwidth,
            height=.205\textwidth,
            symbolic x coords={8, 16, 24, 32},
            xtick style={draw=none},
            xtick distance=1,
            ymax=2,
            ytick distance=0.5,
            x tick label style={yshift={0.5em}},
            xlabel={\small Number of GPUs},
            ylabel={\small Per-iteration time (s)},
            legend image code/.code={ \draw [#1] (0cm,-0.1cm) rectangle (0.2cm,0.15cm); },
            legend style={nodes={scale=0.8, transform shape}, column sep = 10pt, legend columns = -1, legend to name = grouplegend}]

            \addplot[style={fill=orange!60}] coordinates { (8,0.1781228185) (16,0.2582041025) (24,0.3493276834) (32,0.4261697173) };
            \addplot[style={fill=lime!80,postaction={pattern=north east lines}}] coordinates { (8,0.8706123948) (16,1.073997784) (24,1.196673095) (32,1.201170325) };
            \addplot[style={fill=brown!90,postaction={pattern=crosshatch}}] coordinates { (8,0.9846806526) (16,1.096960115) (24,1.213397071) (32,1.312466729) };
            \addplot[style={fill=violet!30,postaction={pattern=horizontal lines}}] coordinates { (8,0.9693546124) (16,1.021469255) (24,1.107851006) (32,1.228989444) };
            \addlegendentry{\OurSystem}
            \addlegendentry{DP-EV}
            \addlegendentry{DeepSpeed}
            \addlegendentry{TAG}
        \end{axis}
        \begin{axis}[name=vit,
            at={($(vgg.north east)+(2em,0)$)}, anchor=north west,
            title = {\small ViT},
            title style={yshift=-.5em},
            ybar, ymin=0, bar width=.12cm,
            enlarge x limits=0.16,
            width=.29\textwidth,
            height=.205\textwidth,
            symbolic x coords={8, 16, 24, 32},
            xtick style={draw=none},
            xtick distance=1,
            ymax=1,
            ytick distance=0.5,
            x tick label style={yshift={0.5em}},
            xlabel={\small Number of GPUs}]

            \addplot[style={fill=orange!60}] coordinates { (8,0.4593142271) (16,0.4632007718) (24,0.5698991179) (32,0.6178576589) };
            \addplot[style={fill=lime!80,postaction={pattern=north east lines}}] coordinates { (8,0.5235230684) (16,0.601987648) (24,0.6196445107) (32,0.7364034772) };
            \addplot[style={fill=brown!90,postaction={pattern=crosshatch}}] coordinates { (8,0.4901064634) (16,0.5506499171) (24,0.6192664504) (32,0.6837266326) };
        \end{axis}
        \begin{axis}[name=bert,
            at={($(vit.north east)+(2em,0)$)}, anchor=north west,
            title = {\small BERT-Base},
            title style={yshift=-.5em},
            ybar, ymin=0, bar width=.095cm,
            enlarge x limits=0.17,
            width=.32\textwidth,
            height=.205\textwidth,
            symbolic x coords={8, 16, 24, 32},
            xtick style={draw=none},
            xtick distance=1,
            ymax=2,
            ytick distance=0.5,
            x tick label style={yshift={0.5em}},
            xlabel={\small Number of GPUs}]

            \addplot[style={fill=orange!60}] coordinates { (8,0.8048781276) (16,0.8935977221) (24,0.9115303755) (32,0.9996345162) };
            \addplot[style={fill=lime!80,postaction={pattern=north east lines}}] coordinates { (8,0.9098040462) (16,1.030650878) (24,1.109551382) (32,1.141639113) };
            \addplot[style={fill=brown!90,postaction={pattern=crosshatch}}] coordinates { (8,1.07868346) (16,1.184060252) (24,1.209952877) (32,1.230797064) };
            \addplot[style={fill=violet!30,postaction={pattern=horizontal lines}}] coordinates { (8,1.165966727) (16,1.263612804) (24,1.500273135) (32,1.639831167) };
        \end{axis}
        \begin{axis}[name=moe,
            at={($(bert.north east)+(2em,0)$)}, anchor=north west,
            title = {\small BERT-MoE},
            title style={yshift=-.5em},
            ybar, ymin=0, bar width=.12cm,
            enlarge x limits=0.13,
            width=.24\textwidth,
            height=.205\textwidth,
            symbolic x coords={8, 16, 24, 32},
            xtick style={draw=none},
            xtick distance=1,
            ymax=4.5,
            ytick distance=1,
            x tick label style={yshift={0.5em}},
            xlabel={\small Number of GPUs}]

            \addplot[style={fill=orange!60}] coordinates { (8,1.492989445) (16,1.962854552) (24,2.461064339) (32,3.032104731) };
            \addplot[style={fill=brown!90,postaction={pattern=crosshatch}}] coordinates { (8,1.810849726) (16,2.158357513) (24,2.784333622) (32,3.167067707) };
        \end{axis}
        \node at ($(7.9cm,-1.2cm)$) {\ref{grouplegend}};
    \end{tikzpicture}

    \vspace{-.8em}
    \caption{Per-iteration training time on homogeneous clusters.\label{fig:exp_homo}}
    \vspace{-.2em}
\end{figure*}
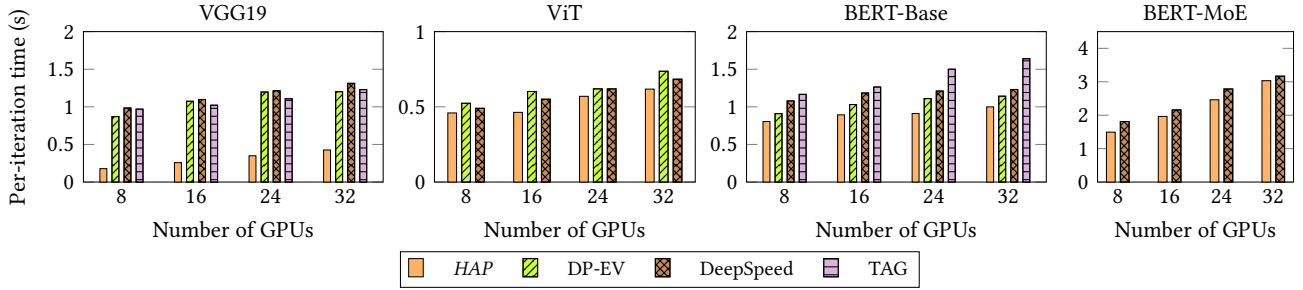

\subsection{Experimental Setup\label{sec:exp_setup}}

\noindent\textbf{Testbed.} We conduct experiments on 8 machines in a public cloud with 64 GPUs in total. Two machines
are each equipped with 8 NVIDIA V100 GPUs and NVLink. The others are each equipped with 8 NVIDIA P100 GPUs.
Inter-machine bandwidth is about 10.4Gbps, as measured with iperf3 \cite{iperf3}. The cluster provides network isolation
and stable bandwidths.

\vspace{1mm}
\noindent\textbf{Benchmarks.} We train 4 representative DNN models as listed in Table \ref{tab:benchmark_models}. VGG19
\cite{vgg} is a convolutional neural network (CNN) for image classification. ViT\cite{vit} is a Transformer-based neural
network for image classification. BERT-Base \cite{bert} is a Transformer-based model for language modeling. Bert-MoE
adds MoE layers to the BERT-Base model by replacing a feed-forward module every two layers in a similar way as in GShard
\cite{gshard}. We follow the convention of scaling MoE models with the number of devices, thus the model size
depends on the number of devices $m$. We adopt weak scaling and set the global batch size proportional to the number of
devices, with per-device batch size 32 for BERT-MoE and 64 for other models.

We use Cifar-10 \cite{cifar} dataset for image classification tasks and WikiText-2 \cite{wikitext} dataset for language
modeling tasks.

\begin{table}[t]
    \vspace{-.5em}
    \caption{Benchmark models\label{tab:benchmark_models}}
    \vspace{-1em}
    \renewcommand{\arraystretch}{1.1}
    \begin{center}
    \begin{small}
    \resizebox{\columnwidth}{!}{
    \begin{tabular}{ | l | l | c | } \hline
        \textbf{Model} & \textbf{Task} & \textbf{Parameters (Millions)} \\ \hline
        VGG19\cite{vgg} & Image Classification & 133 \\ \hline 
        ViT\cite{vit} & Image Classification & 54 \\ \hline 
        BERT-Base\cite{bert} & Language Model & 102 \\ \hline 
        BERT-MoE\cite{bert} & Language Model & 84 + 36m \\ \hline
    \end{tabular}
    }
    \end{small}
    \end{center}
    \vspace{-1em}
\end{table}

\vspace{1mm}
\noindent\textbf{Baselines.} We compare \OurSystem with four relavent designs:
(1) \textit{DP-EV} is data parallelism with even sharding ratios.
(2) \textit{DP-CP} is data parallelism with sharding ratios proportional to the computation speed of the devices. We use
PyTorch's DDP module \cite{ddp} to implement DP-EV and DP-CP.
(3) \textit{DeepSpeed} \cite{deepspeed} supports ZeRO-based \cite{zero} data parallelism and implements intra-op model
parallelism for MoE layers.
(4) \textit{TAG} \cite{tag} is a heterogeneity-aware DNN training system. TAG supports data parallelism and inter-op
model parallelism. It optimizes communication by selecting parameter-server \cite{ps} or \texttt{All-Reduce} for
gradient synchronization and automatically applying sufficient factor broadcasting.

\OurSystem, DP-EV, DP-CP, and DeepSpeed are based on PyTorch and use the same implementation of the benchmark models.
TAG is implemented on TensorFlow \cite{tensorflow}. We were only able to train VGG19 and BERT-Base with TAG. Due to
replicating the whole model on all devices, DP-CP and DP-EV causes out-of-memory errors when training BERT-MoE.

\subsection{Training Speed-up on Heterogeneous Clusters\label{sec:exp_het}}

We first evaluate \OurSystem and the baselines on the heterogeneous cluster with 8 machines. As shown in
Fig.~\ref{fig:exp_het}, \OurSystem significantly outperforms the DP baselines when training VGG19. VGG19 comprises
layers of different computation-to-communication ratios. In particular, the fully-connected layers in VGG19 is very
communication-intensive as compared to the convolution layers. \OurSystem adopts model parallelism to reduce the
communication time and achieves up to 2.41x speedup in the case of 32 GPUs. TAG puts these layers exlusively on one
device in the case of 8 GPUs to eliminate communication. However, this method does not work with more GPUs. \OurSystem
achieves similar performance to DP-CP when training ViT while consistently outperforming the baselines when training
BERT-Base, with a 28\% speedup in the case of 64 GPUs. When training BERT-MoE, \OurSystem finds a strategy that performs
similarly to the expert-designed MoE sharding strategy implemented in DeepSpeed.

\subsection{Training Speed-up on Homogeneous Clusters\label{sec:exp_homo}}

In this experiment, we assess the performance of \OurSystem and baselines on a homogeneous subset of our testbed
consisting of 4 machines, each equipped with 8 P100 GPUs. Since all devices have the same computational power, DP-CP is
equivalent to DP-EV and therefore is not included in this experiment. As demonstrated in Fig.~\ref{fig:exp_homo},
\OurSystem still outperforms all baselines across all models, achieving up to 217\%, 19\%, 22\%, and 13\% speedup when
training VGG19, ViT, BERT-Base, and BERT-MoE, respectively, compared to the best baselines.

\subsection{Ablation Study}

\begin{figure}[t]
    \centering
    \begin{tikzpicture}[ scale=0.9, every node/.style={transform shape} ]
        \begin{axis} [
            ybar stacked,
            ymin=0,
            bar width=.3cm,
            width=.83\columnwidth,
            height=.46\columnwidth,
            symbolic x coords={VGG19, ViT, BERT-Base~~, ~~~~BERT-MoE},
            xtick style={draw=none},
            xtick distance=.98,
            enlarge x limits=0.17,
            ylabel={Throughput (\%)},
            ymin=0,
            ymax=110,
            x tick label style={yshift={0.1em}},
            legend image code/.code={ \draw [#1] (0cm,-0.1cm) rectangle (0.2cm,0.15cm); },
            legend style={nodes={scale=0.85, transform shape}, row sep = .08cm, legend to name = legend},
            reverse legend,
        ]
            \addplot[style={fill=lime!80,postaction={pattern=north east lines}}] coordinates { (VGG19, 41.92960714) (ViT, 82.41972755) (BERT-Base~~, 74.5064823) (~~~~BERT-MoE, 0) };
            \addplot[style={fill=blue!40},postaction={pattern=north west lines}] coordinates { (VGG19, 50.669228) (ViT, 12.3835519) (BERT-Base~~, 11.73962935) (~~~~BERT-MoE, 90.29478134) };
            \addplot[style={fill=orange!60},postaction={pattern=dots}] coordinates { (VGG19, 7.401164857) (ViT, 5.196720546) (BERT-Base~~, 13.75388835) (~~~~BERT-MoE, 9.705218657) };
            \addplot[style={fill=red!60}] coordinates { (VGG19, 0) (ViT, 0) (BERT-Base~~, 0) (~~~~BERT-MoE, 0) };
            \addlegendentry{DP-EV}
            \addlegendentry{Q}
            \addlegendentry{B}
            \addlegendentry{C}
        \end{axis}

        \node at ($(6.3cm,1.4cm)$) {\ref{legend}};
    \end{tikzpicture}
    \vspace{-.5em}
    \caption{Ablation study.\label{fig:exp_ablation}}
    \vspace{-1em}
\end{figure}
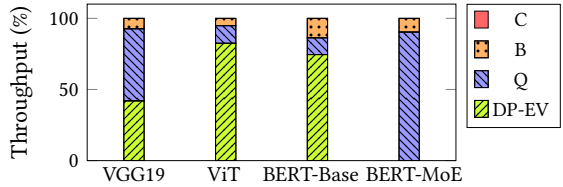

We examine the efficacy of various components of \OurSystem by comparing the throughput of benchmark models achieved
through the utilization of different parts of our designs. In Fig.~\ref{fig:exp_ablation}, DP-EV represents the
throughput achieved without any of our designs. ``Q'' denotes the additional throughput obtained by employing
\OurSystem's program synthesizer. ``B'' represents the throughput contributed by our load balancer, and ``C'' is the
speedup provided by communication optimization. The findings indicate that the program synthesizer has the greatest
impact on the performance of \OurSystem, whereas the communication optimization does not yield noticeable speedup in
this experimental setup. This can be attributed to the relatively small disparity in computational power between the
GPUs. As discussed in Sec.~\ref{sec:motivation_comm}, the communication optimization is mostly effective when there is a
significant difference in sharding ratios between devices.

\subsection{Case Study: Training Multiple Models}

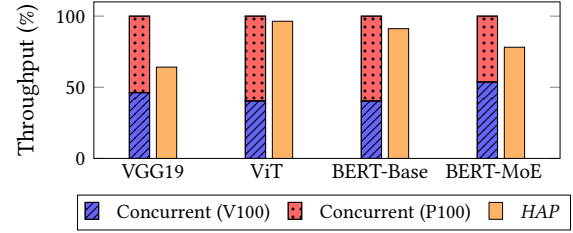
\begin{figure}[t]
    \centering
    \begin{tikzpicture}[
        scale=0.9,
        every node/.style={transform shape},
        every axis/.style={
            ymin=0,
            bar width=.3cm,
            width=1\columnwidth,
            height=.46\columnwidth,
            symbolic x coords={VGG19, ViT, BERT-Base, BERT-MoE},
            xtick style={draw=none},
            xtick distance=.98,
            enlarge x limits=0.17,
            ylabel={Throughput (\%)},
            ymin=0,
            ymax=110,
            xlabel=\phantom{BERT-Base},
            x tick label style={yshift={0.1em}},
        }
    ]
        \begin{axis} [ ybar stacked, bar shift=-.2cm,
            legend image code/.code={ \draw [#1] (0cm,-0.1cm) rectangle (0.2cm,0.15cm); },
            legend style={nodes={scale=0.85, transform shape}, column sep = .2cm, legend columns = -1, legend to name = legend}
        ]
            \addplot[style={fill=blue!60},postaction={pattern=north east lines}] coordinates { (VGG19, 46.17496461) (ViT, 40.32526223) (BERT-Base, 40.24876574) (BERT-MoE, 53.6033766) };
            \addplot[style={fill=red!60},postaction={pattern=dots}] coordinates { (VGG19, 53.82503539) (ViT, 59.67473777) (BERT-Base, 59.75123426) (BERT-MoE, 46.3966234) };
            \addplot[style={fill=orange!60}] coordinates { (VGG19, 0) (ViT, 0) (BERT-Base, 0) (BERT-MoE, 0) };
            \addlegendentry{Concurrent (V100)}
            \addlegendentry{Concurrent (P100)}
            \addlegendentry{\OurSystem}
        \end{axis}

        \begin{axis} [ ybar, bar shift=.2cm, hide axis ]
            \addplot[style={fill=orange!60}] coordinates { (VGG19, 64.16627422) (ViT, 96.39310942) (BERT-Base, 91.1741897) (BERT-MoE, 78.13651425) };
        \end{axis}

        \node at ($(3.4cm,-.8cm)$) {\ref{legend}};
    \end{tikzpicture}
    \vspace{-.5em}
    \caption{Training multiple models.\label{fig:exp_multijob}}
    \vspace{-1em}
\end{figure}

Hardware heterogeneity presents inherent challenges for distributed DNN training. Even with the optimizations of
\OurSystem, it is anticipated that there will be reduced hardware utilization on heterogeneous clusters. We estimate
this overhead by simultaneously training multiple models on homogeneous subsets of the cluster and use the total
throughput as an estimation of the potential throughput achievable if the cluster were homogeneous. Specifically, in
this experiment, we train one model using two V100 machines while simultaneously training another model using six P100
machines. We refer to this approach as \textit{concurrent}. We then compare the total throughput achieved using
\textit{concurrent} with that of \OurSystem, as shown in Fig.~\ref{fig:exp_multijob}. We normalize the throughput of
different models by comparing them to the total throughput achieved by \textit{concurrent}. Our results show that
\OurSystem achieves 64\% to 96\% throughputs compared to \textit{concurrent} for the benchmark models. The VGG19 model
exhibits suboptimal utilization of GPU resources when scaled to accommodate a higher number of devices due to its
relatively small convolution layers. As we scale the MoE model with the number of devices, the BERT-MoE model trained
with \OurSystem is larger than the two models trained using the \textit{concurrent} method.
The results show that \OurSystem facilitates the training of larger models that may exceed the capacity of homogeneous
subsets while maintaining satisfactory throughput on heterogeneous clusters. Given the current trend of large models,
there is a growing demand for the ability to utilize all available resources to train models of maximum size. Moreover,
\OurSystem enables the prioritization of time-sensitive training tasks by fully leveraging all resources for their
execution. For instance, with \OurSystem, users can employ the entire cluster to train a production model before
lower-priority research jobs. On the other hand, \textit{concurrent} limits the training of the production model to a
homogeneous sub-cluster, leading to increased latency.

\subsection{Case Study: Uneven Placement of Experts}

Expert parallelism, which partitions MoE layers on the expert dimension, is the dominating training strategy for MoE
models. Current training systems that adopt expert parallelism allocate the same number of experts to all devices
\cite{gshard,deepspeedmoe,fastmoe}. If the number of devices does not evenly divide the number of experts, the related
tensors must be first padded, resulting in inefficient use of computation power. With uneven partitioning, \OurSystem
naturally supports sharding MoE models with any number of experts onto a cluster with any number of devices, as long as
the total memory capacity is sufficient.

\begin{figure}[t]
    \centering
    \begin{tikzpicture}[scale=0.9, every node/.style={transform shape}]
        \begin{axis}[
            xlabel=Number of experts,
            ylabel=Per-iteration time (s),
            xtick distance=4,
            xmin=3.5,
            xmax=32.5,
            width=.8\linewidth,
            height=.55\linewidth,
            legend style={at={(0.02,0.98)},anchor=north west,nodes={scale=0.85, transform shape}}
        ]

        \addplot[mark=+,blue!60] coordinates { (4,0.2951550961) (5,0.3285242558) (6,0.3291516066) (7,0.3677392403) (8,0.3701153755) (9,0.4117621501) (10,0.429036204) (11,0.4285318851) (12,0.4293793427) (13,0.4308093548) (14,0.4506121476) (15,0.4788458268) (16,0.4985221148) (17,0.5000245333) (18,0.5311600049) (19,0.556898427) (20,0.5604718049) (21,0.6024703185) (22,0.627384154) (23,0.6303825776) (24,0.6527881463) (25,0.668139561) (26,0.6715470314) (27,0.707419014) (28,0.7404125055) (29,0.7484507958) (30,0.7672210455) (31,0.8179775) (32,0.8192022367) };
        \addplot[mark=x,mark repeat=2,brown!90] coordinates { (4,0.3936382294) (4,0.4708968321) (8,0.4708968321) (8,0.5873621941) (12,0.5873621941) (12,0.6449996551) (16,0.6449996551) (16,0.7842837016) (20,0.7842837016) (20,0.9560429811) (24,0.9560429811) (24,1.045114255) (28,1.045114255) (28,1.226110156) (32,1.226110156) };

        \legend{\OurSystem, DeepSpeed}
        \end{axis}
    \end{tikzpicture}
    \vspace{-.8em}
    \caption{BERT-MoE performance with uneven placement of experts.\label{fig:exp_moe}}
    \vspace{-.8em}
\end{figure}
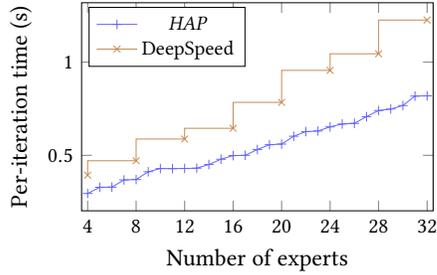

To demonstrate the effectiveness of \OurSystem, we conduct an experiment using two machines, one with 2 NVIDIA A100 GPUs
and the other with 2 NVIDIA P100 GPUs. We train BERT-MoE with varying numbers of experts using \OurSystem and DeepSpeed.
To maintain the same load of each expert, we keep the number of tokens proportional to the number of experts. The
per-iteration training time is plotted in Figure \ref{fig:exp_moe}. DeepSpeed has to pad the number of experts to a
multiplier of 4, the number of available devices, while \OurSystem can provide a smooth performance curve. Further,
\OurSystem places more experts onto A100 GPUs to maximumly exploiting its computation power, bringing up to 64\%
speedup.

\subsection{Cost Model Accuracy}

\OurSystem employs a cost model (Sec.~\ref{sec:cost_model}) to evaluate distributed program $Q$ and sharding ratios $B$
during the optimization loop. The accuracy of the cost model is critical in obtaining the optimal distributed programs.
In this experiment, we alter the configurations (the number of layers, hidden width, and sequence length) of the
BERT-Base model to create different variants and compare the cost estimated by our cost model and the actual profiled
per-iteration training time. As shown in Fig.~\ref{fig:cost_model_accuracy}, the cost model tends to under-estimate the
training time, but the estimated time is mostly linear to the actual time, with a Pearson correlation coefficient of
0.970.

\begin{figure}[t]
    \begin{minipage}[t]{0.47\linewidth}
        \begin{tikzpicture}[scale=0.85, every node/.style={transform shape}]
            \begin{axis}[
                xlabel={Actual time (s)},
                ylabel={Estimated time (s)},
                xmin=-.2, ymin=-.2,
                xmax=2, ymax=2,
                width=1.2\linewidth,
                height=.9\linewidth,
                scatter/classes={a={mark=o,draw=black}},
                y label style={at={(axis description cs:-0.25,1.1)},anchor=east},
            ]
            \addplot[scatter, only marks, mark options={scale=0.5}, blue!80] plot coordinates {
                (1.448296135,1.176223853)
                (1.908091631,1.718327896)
                (0.3860428762,0.3904472912)
                (0.7222995615,0.7638609853)
                (1.232652664,1.078577748)
                (0.2643155336,0.283643224)
                (0.8940812588,0.6856073272)
                (1.420845556,1.006058523)
                (0.3531061459,0.2239305518)
                (0.5564798069,0.4427348616)
                (1.005824442,0.627803077)
                (0.2373094463,0.1622112385)
            };
            \addplot[blue!80,dashed] plot coordinates { (0, 0) (2, 2) };
            \end{axis}
        \end{tikzpicture}
        \vspace{-.8em}
        \caption{Cost model accuracy.\label{fig:cost_model_accuracy}}
        \vspace{-1em}
    \end{minipage}
    \hfill
    \begin{minipage}[t]{0.47\linewidth}
        \begin{tikzpicture}[scale=0.85, every node/.style={transform shape}]
            \begin{axis}[
                xlabel=Number of layers,
                ylabel=Time (s),
                width=1.2\linewidth,
                height=.9\linewidth,
                xmin=0, xmax=25,
                ymin=0, ymax=7.5,
                xtick={0,4,8,12,16,20,24}
            ]
            \addplot[blue!80] plot coordinates { (1,0.06388) (2,0.1102) (3,0.1655) (4,0.2335) (5,0.3071) (6,0.4206) (7,0.5347) (8,0.65) (9,0.8052) (10,1.019) (11,1.226) (12,1.491) (13,1.748) (14,2.069) (15,2.375) (16,2.758) (17,3.162) (18,3.637) (19,4.058) (20,4.55) (21,5.128) (22,5.556) (23,6.173) (24,6.817) };
            \end{axis}
        \end{tikzpicture}
        \vspace{-.8em}
        \caption{Program synthesis time.\label{fig:exp_synthesis_time}}
        \vspace{-1em}
    \end{minipage}
\end{figure}

\subsection{Overhead of \OurSystem\label{sec:exp_synthesis_time}}

\OurSystem adopts SPMD parallelism and exhibits constant search and compiling time with respect to the number of
devices. Therefore, we assess the overhead of \OurSystem by varying model scales. We adjust the number of layers of the
ViT model and generate a distributed model with \OurSystem. The sharding ratio optimization takes less than 1ms and the
majority of overhead is attributed to the program synthesis process. As shown in Fig.~\ref{fig:exp_synthesis_time}, the
program synthesis time increases superlinearly as the number of layers increases. Nevertheless, for a model with 24
layers, \OurSystem only takes a few seconds to synthesize the distributed program, which is negligible compared to the
hours or even days of model training time.

\section{Related Work\label{sec:related_work}}

\noindent\textbf{Inter-Operator Parallelism.}
Placing different parts of the DNN model on different devices allows distributed training on heterogeneous clusters \cite{hdp,placeto,gdp}.
Pure inter-operator parallelism falls short when a single operator in a model exceeds the memory capacity of a single
device, like in the increasingly common MoE models. Inter-operator parallelism may not scale well as each
device is treated individually and the decision space grows
with the number of devices.

\vspace{1mm}
\noindent\textbf{Unevenly-split Data Parallelism.}
VirtualFlow \cite{virtualflow} splits a mini-batch into \textit{virtual nodes} and assigns multiple virtual nodes to a
single device.
HeteroG \cite{heterog} uses graph neural networks and reinforcement learning to find the placement and communication
strategy for each operator in a heterogeneous cluster, supporting both inter-operator parallelism and unevenly-split
data parallelism.
Data parallelism does not support large operations that do not fit in a single device.

\vspace{1mm}
\noindent\textbf{Asynchronous Data-Parallel Training.}
HetPipe \cite{hetpipe} divides the heterogeneous cluster into $k$ virtual workers; each virtual worker
employs pipeline parallelism internally, while asynchronous data parallel training is carried out among virtual workers
using a parameter-server architecture. Prague \cite{prague} adopts partial all-reduce with only a
subset of workers participating in parameter synchronization of each training iteration. Devices with different speeds synchronize
with other devices at different paces.
\OurSystem focuses synchronous training which achieves the same model convergence as single-device training.

\vspace{1mm}
\noindent\textbf{Heterogeneous SPMD Systems.}
AccPar \cite{accpar} uses dynamic programming to decide tensor partitioning among heterogeneous devices, but only considers three types of partitioning for operators in CNN models.
Pathways \cite{pathways}
places model components on different TPU pods and uses gang scheduling for
asynchronously execution of these components.
\OurSystem systematically explores more sharding strategies with program synthesis.

\vspace{1mm}
\noindent\textbf{Collective Communication on Heterogeneous Clusters.} TACCL \cite{taccl} models communication as an
mixed integer linear programming problem and finds routing and scheduling of each data chunk to minimize communication.
BlueConnect \cite{blueconnect} decomposes \texttt{All-Reduce} to fit into heterogeneous network hierarchy. \OurSystem
uses NCCL as the communication library and automatically chooses communication primitives during program synthesis.
\OurSystem may be used together with the heterogeneity-aware communication optimizations to further accelerate training
on heterogeneous clusters.

\section{Conclusion\label{sec:conclusion}}

This paper introduces \OurSystem, an automated system for SPMD-parallel training of large neural networks on
heterogeneous clusters. \OurSystem novelly synthesizes a distributed program on a distributed instruction set that
emulates the single-device program, and identifies the best sharding strategy and communication methods in the
distributed program. Tensor sharding ratios are optimally set to balance the workload across devices, through iterative
optimization with the distributed program synthesis. We implement \OurSystem using PyTorch and demonstrate that it
achieves up to 2.41x faster training compared to existing methods on heterogeneous clusters and can automatically find
feasible SPMD strategies to train large models.

\begin{acks}
This work was supported in part by Alibaba Group through Alibaba Innovative Research (AIR) Program and grants from
Hong Kong RGC under the contracts HKU 17208920 and C7004-22G (CRF).
\end{acks}

\bibliographystyle{ACM-Reference-Format}
\bibliography{main}

\newpage
\appendix
\section{Artifact Appendix}

\subsection{Abstract}

\OurSystem is implemented as a Python module that automatically transforms a single-device tensor program into a
distributed program that can efficiently run on heterogeneous clusters. The artifacts include the source code of
\OurSystem and a Docker container that bundles all dependencies.

\subsection{Description \& Requirements}

\subsubsection{How to access}
The source code of \OurSystem is provided at \url{https://github.com/alibaba/hap}. The ``ae'' branch contains the
version for artifact evaluation. A docker container with all dependencies pre-installed is available at
\url{https://hub.docker.com/r/ylxdzsw/hap}.

\subsubsection{Hardware dependencies}
At least two GPUs are required to run \OurSystem. As \OurSystem is designed for heterogeneous clusters, multiple
machines with different GPU models are needed to fully show \OurSystem's capabilities. The minimum GPU memory should be
at least 12GB. We recommend a similar setting as used in our experiments (Sec.~\ref{sec:exp_setup}), i.e., 2 machines
each equipped with 8 NVIDIA V100 GPUs and 6 machines each equipped with 8 NVIDIA P100 GPUs. The inter-machine bandwidth
is about 10.4Gbps.

\subsubsection{Software dependencies}
\OurSystem is implemented on PyTorch 1.13.1. All machines should use the same versions of CUDA and NVIDIA drivers that
are compatible with PyTorch 1.13.1. Rust 1.70.0-nightly and Coin CBC 2.9.9 are required to build \OurSystem from source.
To reproduce the results of the baselines, DeepSpeed 0.9.4 is also used.

All software dependencies are included and pre-compiled in the Docker image. However, NVIDIA driver 515.43.04 needs to
be separately installed on the host machines.

\subsubsection{Benchmarks}
The benchmark models and datasets are included in the source code repository and Docker image.

\subsection{Set-up\label{sec:artifact_setup}}

This set-up instruction uses the Docker image. To build \OurSystem from source, we refer to the ``readme'' file in the
source code repository.

First, ensure that NVIDIA driver 515.43.04 or higher has been installed on the host machines. The installation can be
verified with the \texttt{nvidia-smi} command. All machines should use the exact same version of NVIDIA driver. The
driver can be installed by following \url{https://docs.nvidia.com/datacenter/tesla/tesla-installation-notes/index.html}.

Next, install Docker engine by following \url{https://docs.docker.com/engine/install}. Then install
\texttt{nvidia-container-toolkit} (e.g., with \texttt{apt-get install}) and restart the docker daemon (\texttt{systemctl
restart docker}). After that, download the Docker image of \OurSystem using \texttt{docker pull ylxdzsw/hap:ae}. The
image is about 20GB. When finished, start a container instance with \texttt{docker run -d -{}-shm-size="10.24gb"
-{}-name hap -{}-gpus all -{}-network host -it ylxdzsw/hap:ae /bin/bash}. To access the container on the host machine,
run \texttt{docker exec -it hap bash}. Inside the container, run \texttt{/usr/sbin/sshd} to start an \texttt{ssh}
instance on port 3922 which will later be used for communication between the containers.

Running HAP involves running the same script on all machines in the cluster. To automate this process, we provide a
helper script \texttt{/root/hap/run\_all}. Running this script on one of the machines starts the same script on all
machines. By default it assumes 8 machines with host names \texttt{v1}, \texttt{v2}, \dots, \texttt{v8}. The IP
addresses of the machines are set in \texttt{/root/.ssh/config}. Before using the script, first enter \texttt{v1} and
run ssh from the \texttt{v1} to all machines (including \texttt{v1} itself) with \texttt{ssh -p 3922 root@vx} and save
the ssh fingerprints. Ensure that \texttt{v1} can access all workers without further interactions such as confirming
fingerprints or typing passwords. When testing HAP on a single machine, one may edit \texttt{/root/hap/run\_all} to keep
only the line with \texttt{v1} and edit \texttt{/root/.ssh/config} to set the ip address of \texttt{v1} to 127.0.0.1.

Finally, check the set-up by running \texttt{./run\_all worker.py 1} on the \texttt{v1}. It should run 100 iterations of
training and reports the average per-iteration time.

\subsection{Evaluation workflow\protect\footnotemark}
\footnotetext{Submission, reviewing and badging methodology followed for the evaluation of this artifact can be found at
\url{https://sysartifacts.github.io/eurosys2024/}.}

\subsubsection{Major Claims}

\begin{itemize}
    \item \textit{(C1)}: \OurSystem outperforms the baselines in heterogeneous clusters in terms of the per-iteration
    training time when training the benchmark models. This is proven in Sec.~\ref{sec:exp_het} and the results are
    shown in Fig.~\ref{fig:exp_het}.

    \item \textit{(C2)}: \OurSystem outperforms the baselines in homogeneous clusters in terms of the per-iteration
    training time when training the benchmark models. This is proven in Sec.~\ref{sec:exp_homo} and the results are
    shown in Fig.~\ref{fig:exp_homo}.

    \item \textit{(C3)}: \OurSystem can generate distributed models within seconds for the benchmark models, as shown
    in Sec.~\ref{sec:exp_synthesis_time} and Fig.~\ref{fig:exp_synthesis_time}.
\end{itemize}

\subsubsection{Experiments}

\textit{Experiment (E1): [Heterogeneous Cluster] [30 human-minutes + 4 compute-hours]}: Train the benchmark models on
a heterogeneous cluster and compare the per-iteration training time of \OurSystem and the baseline systems.

\vskip 1em
\noindent\textit{[Preparation]}

Assuming that \OurSystem has been set up on a heterogeneous cluster following Sec.~\ref{sec:artifact_setup}, this
experiments involves modifying \texttt{config.py} and running \OurSystem and the baselines.

First, we need to collect profiling data. The device flops can be profiled by running \texttt{python profiler.py}.
Execute this command for each type of GPU and replace \texttt{device\_flops} in \texttt{worker.py} with the actual
profiling data. \texttt{device\_flops} is an array of the flops for all devices. For example, when using 2 V100 GPUs and
6 P100 GPUs, it should be set to an array of 8 elements, with the first two elements being the profiled flops of the
V100 GPU and the last 6 elements being the profiled flops of the P100 GPU. The collective communication can be profiled
by running \texttt{./run\_all profiler.py 8}, which automatically runs different collective operators across all
machines using 8 GPUs (the second argument to the script) on each machine. Fill the profiling data in
\texttt{worker.py}.

Next, modify \texttt{config.py} and run \texttt{./run\_all worker.py k} to obtain the per-iteration training time of
\OurSystem, where \texttt{k} is the number of GPUs to use on each machine. In \texttt{config.py}, \texttt{model\_name}
is the benchmark model, where \texttt{Vvgg}, \texttt{Vtransformer}, \texttt{Rtransformer} and \texttt{Rmoe} correspond
to the VGG19, ViT, BERT-Base, and BERT-MoE models. \texttt{world\_size} is the total number of GPUs.
\texttt{master\_addr} should be set to the ip address of one of the machines. \texttt{cards\_per\_node} is only used by
the DeepSpeed baseline and should be set to the number of GPUs to use on each machine (same as \texttt{k}). Other fields
should be kept unchanged to reproduce the results reported in the paper.

To run the DP-EV baseline, change the \texttt{unscaled\_sharding\_lengths} in \texttt{ddp.py} to an array of 1
(simulating the same device flops on each device regardless of their actual types) and run \texttt{./run\_all ddp.py k}
similar to running \OurSystem. To run the DP-CP baseline, fill \texttt{unscaled\_sharding\_lengths} with the actual
profiling data of each GPU type in the same way as \texttt{device\_flops} in \texttt{worker.py}.

To run the DeepSpeed baseline, use \texttt{./run\_all\_deepspeed} instead of \texttt{./run\_all}.

\vskip 1em
\noindent\textit{[Execution]}

To collect the data for Fig.~\ref{fig:exp_het}, vary \texttt{k} and the related fields in \texttt{config.py}
(\texttt{model\_name}, \texttt{world\_size}, and \texttt{cards\_per\_node}), then run \OurSystem and the baselines for
each configuration.

\vskip 1em
\noindent\textit{[Results]}

The experiment scripts print the average per-iteration time and the standard deviation on screen. As the standard
deviation is relatively small in our experiments, we only report the average per-iteration time in Fig.~12. The
experiment script also records the timeline in \texttt{trace.json.gz}, which can be load into Chrome Trace Profiling
Tool for further inspection. The results should confirm the claim C1.

\vskip 1em
\textit{Experiment (E2): [Homogeneous Cluster] [30 human-minutes + 4 compute-hours]}: Train the benchmark models on
a homogeneous cluster and compare the per-iteration training time of \OurSystem and the baseline systems.

\vskip 1em
\noindent\textit{[Preparation]}

The preparation is same as in E1, except for that we now use a homogeneous cluster.

\vskip 1em
\noindent\textit{[Execution]}

Same as in E1.

\vskip 1em
\noindent\textit{[Results]}

Same as in E1. The results should confirm the claim C2.

\vskip 1em
\textit{Experiment (E3): [Overhead] [5 human-minutes + 5 compute-minutes]}: Evaluate the time required by \OurSystem to
generate a distributed program.

\vskip 1em
\noindent\textit{[Preparation]}

This experiment requires only one machine and can run without GPUs. Set \texttt{model\_name} in \texttt{config.py} to
\texttt{Vtransformer} for the ViT model and vary the \texttt{nlayers} field to experiment with models of different
number of layers.

\vskip 1em
\noindent\textit{[Execution]}

Run \texttt{python master.py}. This script compiles the model without actually running it.

\vskip 1em
\noindent\textit{[Results]}

The compile time is printed on the screen. The results should confirm the claim C3.

\subsection{Notes on Reusability\label{sec:artifact_reuse}}

\OurSystem can be extended to support new operators and custom semantic rules. To add support for a new operator, one
need to edit the file \texttt{hap.rs} and add a new handler in the function \texttt{initialize\_parsing\_handlers}
following the same structure of the existing handlers. To add new rules for generating Hoare triples, edit the
\texttt{analyze\_rgraph} function in \texttt{hap.rs}. The existing rules in the code can be used as examples.

\end{document}